\documentclass[journal]{vgtc}                     

\usepackage{tabu}                      
\usepackage{booktabs}                  
\usepackage{lipsum}                    
\usepackage{mwe}                       

\usepackage{mathptmx}                  

\usepackage{amsmath,amsfonts}

\usepackage{algorithm}
\usepackage{algorithmicx}
\usepackage[noend]{algpseudocode}
\usepackage{array}
\usepackage[caption=false,font=normalsize,labelfont=sf,textfont=sf]{subfig}
\usepackage{textcomp}
\usepackage{stfloats}
\usepackage{url}
\usepackage{verbatim}
\usepackage{graphicx}
\usepackage{cite}
\usepackage{bm}
\usepackage{color}
\usepackage{tabu}                      
\usepackage{booktabs}                  
\usepackage[normalem]{ulem}
\usepackage{hhline}
\usepackage{colortbl}
\usepackage{multirow}
\usepackage{mathptmx}                  

\onlineid{1060}

\vgtccategory{Research}

\vgtcpapertype{please specify}

\title{DiffPC: Diffusion-Based Projector Photometric Compensation}

\author{%
  \authororcid{Yuxi Wang}{0000-0001-9358-5893},
  \authororcid{Haibin Ling}{0000-0003-4094-8413}, 
  \authororcid{Bingyao Huang}{0000-0002-8647-5730}
}

\authorfooter{
  \item
  	Yuxi Wang is with Hangzhou Dianzi University.
  	E-mail: yxwang@hdu.edu.cn
  \item 
    Haibin Ling is with Westlake University.
  	E-mail: linghaibin@westlake.edu.cn. Corresponding author.
  \item 
    Bingyao Huang is with Southwest University.
  	E-mail: bhuang@swu.edu.cn. Corresponding author.

}

\abstract{
    Projector photometric compensation corrects color distortions introduced by surface texture, reflection, and ambient lighting. Existing deep learning-based methods usually require professional scene-specific data collection and lack consideration for perceptual quality.
    To address this limitation, we present a diffusion-based photometric compensation method that reconstructs compensation images under photometric and content-aware guidance. Specifically, we first model the photometric distortions introduced during projection as environment-dependent additive noise, thereby reformulating the photometric compensation problem as a denoising task with physical constraints. Next, we introduce a diffusion model, which generates compensation images by following an additive trajectory to iteratively remove the noise. Finally, to accurately estimate the noise at each timestep, 
    by analyzing the factors that contribute to distortions in the physical process of projection and capturing, we design a noise estimation network that incorporates features of both photometry-aware and content conditions.
   Experiments show that our method achieves superior visual performance in unknown scenarios, thereby exhibiting significant practical advantages over prior methods.
    
} 


 \teaser{
    \includegraphics[width=1.0\linewidth]{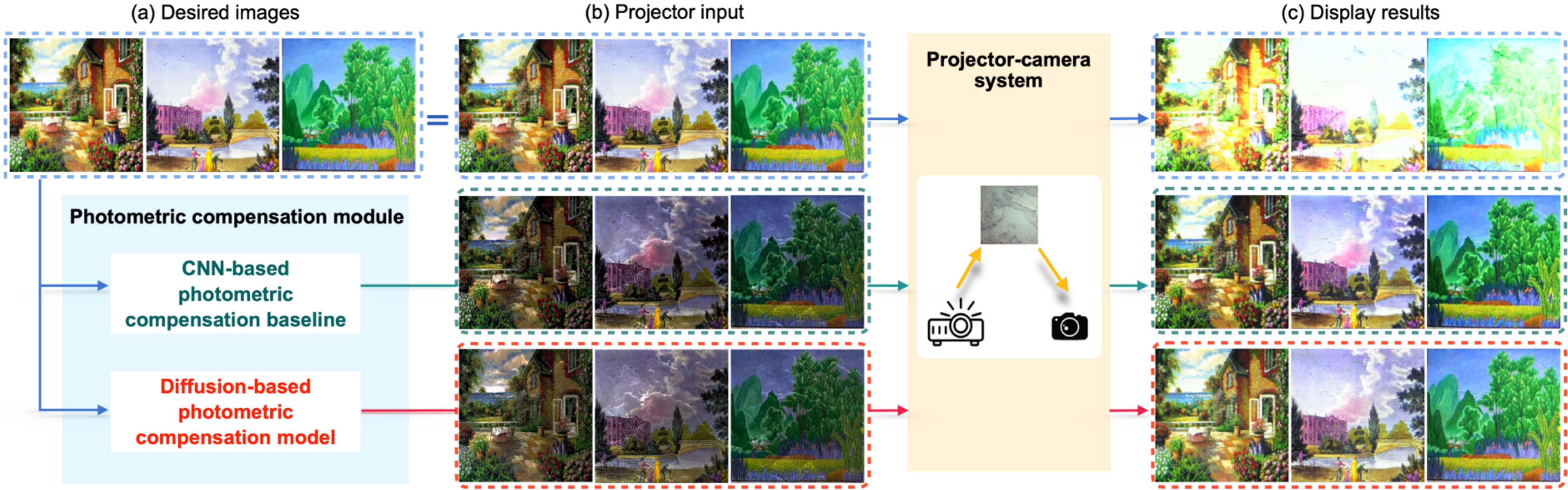}
    \caption{Pipeline of projector photometric compensation and display. The desired images (a) are first processed by the compensation models to generate compensation images, which serve as the projector input (b). Through the projector–camera system, these images are rendered in the environment, and the resulting display results (c) are then captured and observed. Compared with the CNN-based baseline CompenNeSt, the proposed diffusion-based approach enhances visual quality and preserves finer image details. 
}
    \label{fig:teaser}
 }

\begin{document}


\firstsection{Introduction}

\maketitle
    Projectors are widely used display devices in a broad range of applications~\cite{Hamasaki_2018, Miyashita_2018, Ueda_2020, Miyatake_2023, Hiratani_tvcg_2023, Huang_2025_OLE,Yoshida_tvcg_2025}. However, when projecting onto textured surfaces, color distortions caused by surface patterns, reflections, and ambient illumination often result in significant deviations from the intended visual appearance. To address this issue, numerous projector compensation techniques have been proposed ~\cite{ Fujii_CVPR_2005, Narita_TVCG_2017,Shih_TIP_2021, Sugimoto_TVCG_2021, Huang_PAMI_2021, Wang_VR_2023, Li_TVCG_2023, Iwai_review_2024}. 
     
    Traditional approaches are derived based on the radiometric transfer model~\cite{Fujii_CVPR_2005, Bokaris_ECCV_2014}. They estimate the color transfer function by projecting predefined patterns onto the target surface, and then compute pixel-wise compensation for color and brightness. However, these methods are setup-dependent and sensitive to environmental variations. 
    In recent years, deep learning-based methods have been proposed to implicitly model the color transfer function~\cite{Huang_CVPR_2019,Li_TVCG_2023,Wang_TVCG_2024}. After training using scene-specific image pairs, these methods achieve higher compensation accuracy than model-based techniques. Their reliance on extensive data collection and environment-specific training significantly restricts their adaptability in unknown scenarios.
    
    Recently, diffusion-based methods have achieved remarkable progress in image restoration tasks such as deblurring~\cite{Whang_CVPR_2022,Lv_CVPR_2024}, super-resolution~\cite{Saharia_PAMI_2023,Luo_ICML_2023,Wang_CVPR_2024,Zhang_CVPR_2025}, and colorization~\cite{Cong_CVPR_2024,Yan_WACV_2025,Zabari_SIGGRAPH_2023}. These methods restore high-quality images by denoising corrupted input under conditional guidance iteratively, achieving impressive visual perception performance and strong generalization capabilities.
    
    {Inspired by the remarkable success of diffusion models in image restoration tasks, this paper proposes the Diffusion-based Photometric Compensation (DiffPC) method via formulating the projection and capturing process as a specialized image degradation process. 
    In our framework, photometric compensation can be treated as a conditional generation task, where the ideal projector input is recovered from its degraded observation. }    

    Unlike image restoration tasks {that deal with explicit degradation models like additive noise or blurring kernels,} the degradation in our task arises from the nonlinear interaction between global illumination, surface texture, reflectance properties, and the intrinsic characteristics of the hardware devices.
    {To address these complexities, and building upon the insights from \cite{Huang_CVPR_2019} presenting implicit physical modeling, we design a dual-conditioned UNet-based noise estimation network: the desired image provides the visual content, and the captured images of the projection surface serve as photometry-aware guidance.}
    In conjunction with the diffusion timesteps, these conditions and their interactions enable the noise estimation network to iteratively approximate the underlying noise distribution, achieving superior perceptual compensation performance.

    {Leveraging this diffusion-based framework, our DiffPC can effectively generate compensation images in unknown static environments without requiring additional calibration procedures.}
    Furthermore, by fine-tuning the model with a small amount of scene-specific data, substantial performance improvements can be achieved. This capability not only enhances the robustness and adaptability of projector–camera systems but also holds significant potential for their widespread deployment in real-world scenarios.

    Our contributions are summarized as follows:
	\begin{itemize}
        \item {We reformulate the projector photometric compensation task as a conditional generative process, and then present a diffusion-based photometric compensation method to address color distortions caused by complex surface textures and ambient lighting conditions.}
        \item{{We design a dual-conditioned noise estimation network inspired by the physical degradation process. By integrating the noisy state, diffusion timesteps, target content, and photometry-aware surface guidance, the network can iteratively and accurately estimate the noise distribution.}}

        \item {{Our DiffPC exhibits strong generalization capabilities, allowing for direct deployment in unknown scenarios without additional calibration. Moreover, it supports efficient scene-specific adaptation through minimal fine-tuning, significantly enhancing its practicality.}}
        
        \item {Extensive experiments demonstrate that our approach achieves high-quality display results that better align with human visual perception.}
	\end{itemize}

    \section{Related Work}
    In this section, we discuss the related work concerning photometric compensation, diffusion models for image restoration, and projection mapping using diffusion models. 
    \subsection{Photometric compensation}
    Photometric compensation is a key component of projector-camera systems. It involves adjusting the projector input to account for the diverse colors and textures of projection surfaces, as well as the complex lighting conditions in the scene. 
    
    Traditional photometric compensation relies on radiometric transformation models, where the response of each pixel in the captured image can be computed by integrating the product of the projector’s spectral output, the surface reflectance, and the camera’s spectral sensitivity over the visible spectrum\cite{Grossberg_cvpr_2004,Bimber_sig_2008}. Building on this, Nayar \emph{et al}.\cite{Nayar_ICCVW_2003} simplified this projection and capturing process to a linear transformation function by modeling the couplings between the projector and camera channels and their interactions with spectral reflectance using a $3\times3$ color mixing matrix. Grossberg \emph{et al}.\cite{Grossberg_cvpr_2004} extended this model by incorporating the effect of environmental lighting in the scene, and then efficiently calibrated its parameters using only six images. 
    Considering the limitation of the projector's dynamic range, 
    Ashdown \emph{et al}.\cite{Ashdown_CVPRW_2006} integrated this model with image content considerations and adjusted the compensation image according to the limited chrominance and luminance values of the projector. 
   Huang \emph{et al}. \cite{Huang_tip_2017} further refined brightness and hue adjustments based on the chromatic adaptation of the human visual system to reduce color clipping artifacts.  Akiyama \emph{et al}.\cite{Akiyama_VRW_2022} enhanced the visual quality of the projected content via perceptually-based tone mapping optimization. As the color mixing and device responses are complex and cannot be accurately represented by simple linear transformations, {Grundhöfer \emph{et al}.\cite{Grundhofer_cvpr_2013,Grundhöfer_tip_2015} estimated the per-pixel non-linear color mapping using a thin-plate spline, while Kurth \emph{et al}.\cite{Kurth_ISMAR_2020} employed a second-order polynomial to model the nonlinear projector’s response function and cross-channel interference among the three color channels.
   Although these methods achieve good performance in static scenes, they fail to adapt to changes in the projection surface or lighting conditions in dynamic environments. To address these limitations, Fujii \emph{et al}.\cite{Fujii_CVPR_2005} combined physical models with dynamic feedback. It adjusts the reflectance matrix according to the fraction of the captured image and the desired effect when significant changes occur. Amano \emph{et al}.\cite{Amano_ICCV_2010} applied model predictive control (MPC) to iteratively refine the projector input based on the difference between the captured feedback image and the one predicted by the projector camera response model. For a continuously changing non-rigid surface, Hashimoto \emph{et al}. \cite{Hashimoto_TVC_2021} estimated inter-pixel correspondences between the projector and camera in real-time and continuously updated the surface reflectance by minimizing the differences between the calculated and measured values. Additionally, considering the effect of inter-pixel coupling, Shih \emph{et al}. \cite{Shih_TIP_2021} calibrated the gamma function and the inter-channel coupling matrix offline, and then estimated the dynamic reflectance through projecting calibration patterns. This approach offers enhanced accuracy in compensating for surface variations and photometric distortions.
    
    With the remarkable progress of deep learning in computer vision, Huang \emph{et al}. \cite{Huang_CVPR_2019,Huang_PAMI_2021} learned the photometric transformation function directly from pairs of natural projection and captured images using deep neural networks. These approaches demonstrate the effectiveness of deep learning in addressing photometric distortion. 
    Wang \emph{et al}\cite{Wang_VR_2023} further improved the efficiency of \cite{Huang_PAMI_2021} for high-resolution projector-camera systems, enhancing practicality for high-resolution applications. Kageyama \emph{et al}.\cite{Kageyama_OE_2020} presented a deep neural network to compensate for projector blur on textureless surfaces, and then, they\cite{Kageyama_TVCG_2022} extended this work to apply to dynamic projector mapping by leveraging information from the previous frame. 
    Furthermore, Li \emph{et al}.\cite{Li_TVCG_2023} decomposed the spectral power distribution (SPD), and alternately optimized the color mixture matrix and the non-linear projector transfer function modeled by a convolutional neural network. Although these trainable methods exhibit remarkable performance in generating high-quality compensation images, the offline data acquisition process remains complex for non-experts. To overcome this, Wang \emph{et al}.\cite{Wang_TVCG_2024} developed an online video compensation system that, through collaborative multi-threading, automatically performs online compensation, data acquisition, and model updating. This system can be quickly configured and utilized in unknown environments. These deep learning-based methods achieve excellent performance in photometric compensation. However, most of them are highly setup-dependent: models trained under one configuration often generalize poorly to new environments. Moreover, acquiring the large amounts of data required for training is labor-intensive and technically demanding, and the training process itself is computationally expensive. In addition, most methods focus primarily on pixel-level similarity and do not explicitly consider human visual perception, which is essential for achieving high subjective visual quality.
     
    \subsection{Diffusion models for image restoration}
    Image restoration tasks, such as super resolution, image deblurring, inpainting, et al., aim to reconstruct a clean image from its degraded version. Recently, Diffusion models\cite{Ho_NEURIPS_2020,Song_2021_ICLR_scorebased}, which iteratively refine data through a learned denoising process, have been widely applied to image restoration tasks and achieved outstanding performance\cite{Rombach_2022_CVPR,Saharia_PAMI_2023,Chung_2022_CVPR,Kawar_nips_2022,Lv_CVPR_2024,Xiao_TGRS_2024,Luo_2025_aaai}. Existing diffusion-based image restoration methods can be categorized into supervised and zero-shot methods. Saharia\emph{et al.} \cite{Saharia_PAMI_2023} introduces the condition into the diffusion model directly by concatenating the low-quality image with the noisy state. Based on Stochastic Differential Equations (SDEs)\cite{Song_2021_ICLR_scorebased}, Luo \emph{et al.}\cite{Luo_ICML_2023} proposed a unified framework for image restoration that can adapt to various degradation tasks. They applied the mean of the SDE to the degraded image and constructed a closed-form SDE. Whang \emph{et al.}\cite{Whang_CVPR_2022} reduced inference cost and improved the perceptual quality by integrating a deterministic predictor with a stochastic diffusion-based residual refiner. Yue \emph{et al.} \cite{Yue_NIPS_2023} introduced a Markov chain that progressively injects the residual between the high-resolution and low-resolution images, enabling the model to start from a more informative prior. This design significantly reduces the required sampling steps (e.g., to 15–20) while maintaining or even improving image quality. Zhang \emph{et al.}\cite{Zhang_2025_CVPR} introduced a region-wise noise injection mechanism guided by pixel-level uncertainty.  The predicted reconstruction residual of each pixel is used to estimate uncertainty, which in turn adaptively modulates the noise intensity at each diffusion timestep. This strategy improves the preservation of fine-grained local structures in the generated high-resolution images. Ju \emph{et al.}\cite{Ju_eccv_2024} proposed a dual-branch architecture for image inpainting. Features from masked regions are extracted via a dedicated branch and then injected into a pre-trained UNet through zero convolution blocks. By decoupling the feature extraction and generation processes, this approach enhances semantic coherence and pixel-level consistency in inpainted results. Although existing diffusion models make significant progress in restoring degraded images, the restoration of images degraded under physically constrained processes, such as the generation of photometric distortions caused by varying lighting conditions and surface reflectance in a projector-camera system, remains underexplored.

    \subsection{Projection mapping using diffusion model}
     Recently, with the development of Generative AI, 
     several works have incorporated diffusion models into projection mapping.
     Erel \emph{et al}.\cite{Erel_TVCG_2023} proposed a neural reflectance field comprising three neural networks to estimate geometry, material properties, and transmittance from freely captured images. Benefiting from the use of this neural reflectance field, this method can render content generated by a diffusion model from the user-defined viewpoints. Additionally, Erel \emph{et al}. \cite{Erel_SIGGRAPH_2024} presented a cascaded perceptual dynamic projection mapping system for human hands. This system generates visual content using a diffusion model guided by the hand mask and semantic prompt, and then dynamically projects the generated image onto the hand to enhance expressive augmented reality interactions.
     Deng \emph{et al}.\cite{Deng_TVCG_2025} proposed a language-guided style transfer method to reduce the surface texture-related artifacts. While these methods demonstrate the potential of generative models for visual content generation in projection systems, they largely overlook the diffusion process itself and its integration with the physical modeling of projection compensation. 

    \subsection{Our method}
    In this paper, we address the problem of photometric compensation for projector–camera systems operating on textured surfaces under unknown ambient lighting, assuming geometric distortions have already been corrected. We propose a generative diffusion-based framework with improved human perceptual visual quality. 
    Unlike existing CNN-based methods that directly predict the compensation image from the desired effect, {we interpret the discrepancy between the captured observation and the ideal projection as a high-dimensional residual, thereby reformulating the photometric compensation task as a progressive denoising process. By adopting a diffusion-based model in our framework, we iteratively restore the compensation image from its distorted state. This iterative refinement process can effectively handle complex nonlinear projection problems.} Furthermore, since existing diffusion-based image restoration methods often lack a physical prior, we design a specialized noise estimation network. This network jointly integrates photometry-aware and content conditions to ensure accurate noise estimation at each timestep. Consequently, our method exhibits strong generalization ability and can be directly applied to unknown environments with good display image quality and human visual perception performance.


    \section{Preliminaries: Mean-Reverting SDE}
    We correct photometric distortions in {the} projector-camera system using a diffusion-based model. {Specifically}, we adopt IR-SDE\cite{Luo_ICML_2023} as our backbone. The forward process of IR-SDE is formulated as:
    \begin{equation}
        dx=  \theta_t(\mu-x)dt+\sigma_tdw
        \label{eq:irsde-forward}
    \end{equation}
    where $\mu$ denotes the state mean, and $\theta_t$, $\sigma_t$ are time-dependent positive parameters, controlling the speed of the mean-reversion and the stochastic volatility, respectively. {$w$ denotes a standard Wiener process.}
    
    Let the SDE coefficients in \cref{eq:irsde-forward} satisfy $\sigma_t^2/\theta = 2\lambda^2$, {where $\lambda^2$ is the stationary variance.} Then, given any starting state $x(0)$, the transition kernel $p_t$ can be defined as a Gaussian distribution with mean $m_{t}$ and variance $v_{t}$.
    
    In the forward process, the starting state $x(0)$ is degraded to the counterpart $\mu$ by diffusing $x(0)$ towards a noisy version. By reparameterizing, $x(t)$ can be obtained by $x(t)=m_t(x)+\sqrt{v_t}\epsilon_t$, 
    where $\epsilon_t\sim \mathcal{N}(0,I)$ is a standard Gaussian noise.
    
    Subsequently, to recover the initial state $x(0)$ from the terminal state $x(T)$, the reverse-time SDE can be formulated by: 
    \begin{equation}
        dx=  [\theta_t(\mu-x)-\sigma^2\nabla_x\log p_t(x)]dt+\sigma_td\hat{w}
        \label{eq:irsde-reverse}
    \end{equation}
    
     By estimating the noise using a CNN-based network, the ground truth and predicted scores in \cref{eq:irsde-reverse} can be obtained according to $p_t$ and the predicted noise $\hat{\epsilon}_t$ respectively:
    \begin{equation}
        \nabla_x \log p_t(x|x(0))= -\frac{x(t)-m_t(x)}{v_t} = -\frac{\epsilon_t}{\sqrt{v_t}}
        \label{eq:score}
    \end{equation}
    Using this reverse process, the initial state $x(0)$ can be recovered from its degraded state $x(T)$ iteratively.

    Different from other diffusion-based models that learn the score by minimizing the difference between the network output and $\epsilon_t$, \cite{Luo_ICML_2023} provides a new maximum likelihood objective to find more accurate images rather than scores. The objective function can be defined as:
    \begin{equation}
    \begin{split}
    \mathcal{L}_\gamma(\phi) := &\sum_{i=1}^T{\gamma_i\mathbb{E}[\|\hat{x}_{i-1}-x_{i-1}^*\|]}\\
    = &\sum_{i=1}^T{\gamma_i\mathbb{E}[\|x_i-(dx_i)_{\tilde{\epsilon}_\phi(x_i,\mu,i)}-x_{i-1}^*\|]}
    \end{split}
    \label{eq:loss-err-x}
    \end{equation}
    where $\hat{x}_{i-1}$ is the predicted $i-1$ th state that is calculated by $x_i-(dx_i)_{\tilde{\epsilon}_\phi(x_i,\mu,i)}$. $x^*_{i-1}$ is the optimized reverse state. More details can be found in \cite{Luo_ICML_2023}.
    
    \begin{figure*}[!htb]
        \centering
        \includegraphics[width=1.0\linewidth]{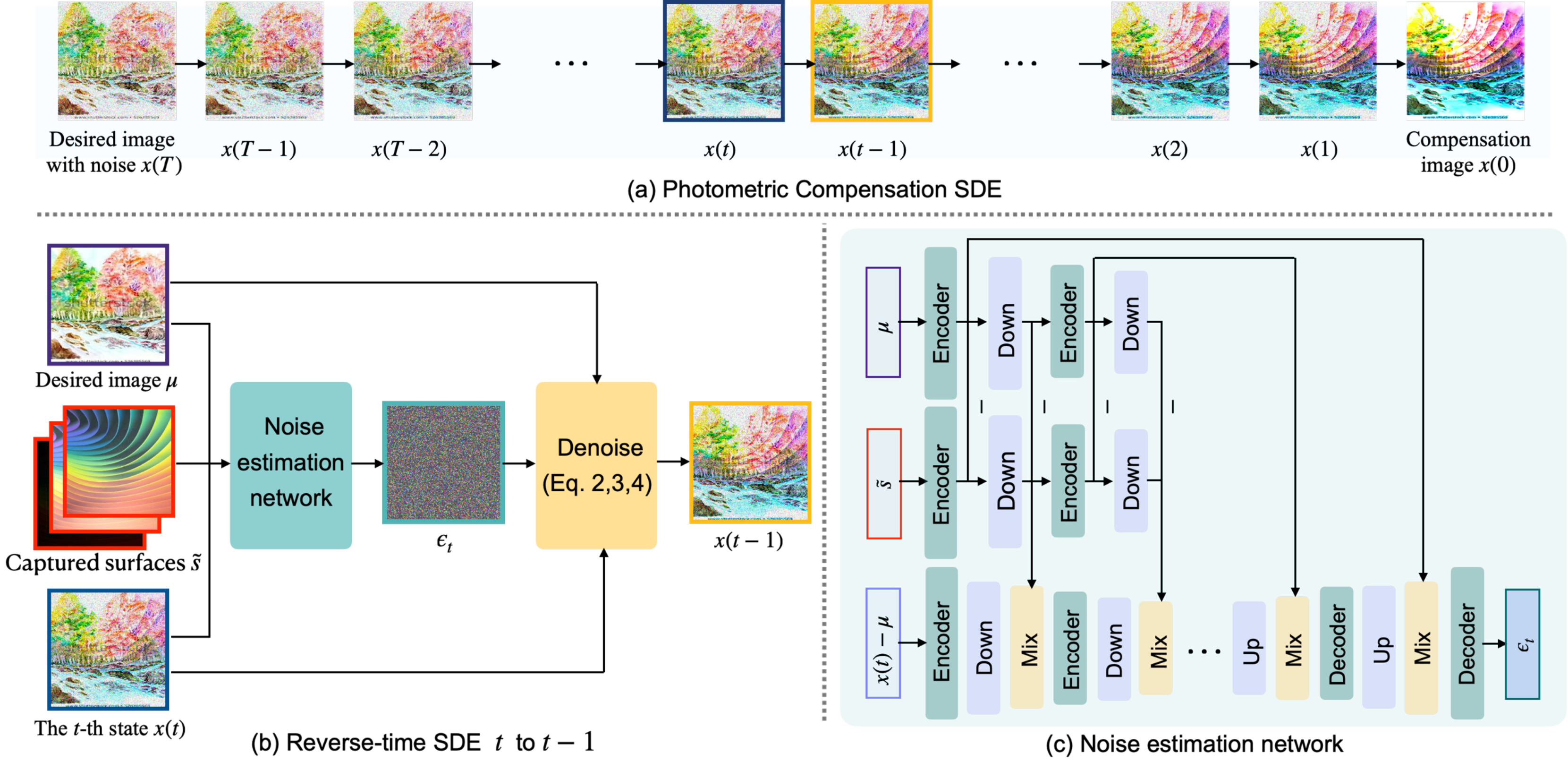}
        \caption{Overview of our diffusion-based photometric compensation framework. (a) The terminal state $x(T)$ is obtained by perturbing the desired image with Gaussian noise. Through the iterative reverse-time SDE, the noise is gradually removed, and the trajectory progressively converges to the initial state $x(0)$, which corresponds to the compensation image. (b) During the reverse-time process at each timestep, a deep neural network is employed to predict the noise. Then, according to \cref{eq:irsde-reverse}, \cref{eq:score}, \cref{eq:loss-err-x},
        the noise score is used to estimate $dx$, which in turn allows the computation of $x(t-1)$. (c) Our noise estimation network incorporates two additional branches to encode conditional features at multiple scales, which are injected into the main branch to perform condition-guided noise estimation, ensuring consistency with the physical process of projection.}
        \label{fig:flowchart}
    \end{figure*}

    \section{Photometric compensation using diffusion model}
    In this section, we first formulate {the} photometric compensation task using a diffusion model, and then design a neural network to estimate the noise in each timestep. The overview of our compensation framework is illustrated in \cref{fig:flowchart}. 
    
    \subsection{Task formulation}
    In the projector-camera system, denote the projector input and the captured image be $\bm{y}$ and $\tilde{\bm{y}}$, the global lighting irradiance distribution be $\bm{g}$, the surface spectral reflectance property be $\bm{s}$, the geometric projection, the radiometric transfer function and spectral reflectance functions be $\mathbb{\pi}_p$ $\mathbb{\pi}_c$, and $\mathbb{\pi}_s$, then the projection and capturing process can be formulated by:
    \begin{equation}
    \begin{split}
    \tilde{\bm{y}}&=\mathbb{\pi}_c(\mathbb{\pi}_s(\mathbb{\pi}_p(\bm{y}),\bm{g},\bm{s}))\\
    \end{split}
    \label{eq:pcs}
    \end{equation}
    
    Define $\mathbb{\pi}_r$ as the composite radiometric transfer function. By employing captured surface image $\tilde{\bm{s}}$ under the global lighting and the projector backlight to capture the spectral interactions of $\bm{g}$ and $\bm{s}$, \cref{eq:pcs} can be rewritten as:
    \begin{equation}
    \begin{split}
    \tilde{\bm{y}}&=\mathbb{\pi}_r(\mathbb{\pi}_p(\bm{y};\tilde{\bm{s}}))\doteq \mathbb{\pi}(\bm{y};\tilde{\bm{s}})\\
    \end{split}
    \label{eq:pcs2}
    \end{equation}
    where $\mathbb{\pi}$ is defined as the function that integrates both geometric and photometric transformations.
    
    Projector compensation aims to find the compensation image $\bm{y}^*$ of $\bm{y}$, whose display effect is as close as possible to the desired image $\bm{y}$. Denote $\mathbb{\pi}^{\dagger}$ be the inverse transformation of $\mathbb{\pi}$, then the compensation and projecting-capturing process can be formulated by the two equations below, respectively:
    \begin{equation}
    \begin{split}
    \bm{y}^* & =  \mathbb{\pi}^{\dagger}(\bm{y};\tilde{{\bm{s}}})\\
    \tilde{\bm{y}}^*&=\mathbb{\pi}(\bm{y}^*;\tilde{\bm{s}})\approx \bm{y}\\
    \end{split}
    \label{eq:full_cmp}
    \end{equation}
    
    Omitting geometric transformations, photometric compensation is formulated as an estimation problem for $\mathbb{\pi}^\dagger$ according to image pairs $(\bm{y},\bm{y}^*)$ and $\tilde{\bm{s}}$. More details can be found in \cite{Huang_PAMI_2021}.

    \subsection{Motivation of our method}
    {As the non-linear interactions between surface geometry, material properties, and lighting conditions are difficult to model explicitly, the color shifts between the projector input and its observation can be effectively reformulated as physical constrained additive residual term.} Thus, \cref{eq:full_cmp} can be rewritten as:
    \begin{equation}
    \begin{split}
        \tilde{\bm{y}}^*=\mathbb{\pi}(\bm{y}^*;\tilde{\bm{s}})&:=\bm{y}^*+\eta\approx \bm{y} \\
        \bm{y}^*  =  \mathbb{\pi}^{\dagger}(\bm{y};\tilde{{\bm{s}}})& =\bm{y}-\eta, 
    \end{split}
    \label{eq:cmp_noise}
    \end{equation}
    where $\eta\sim\mathcal{N}(m(\bm{y},\bm{\tilde{s}}),v(\bm{y,\bm{\tilde{s}}}))$ models surface reflectance effects, ambient lighting, and sensor noise as condition-dependent Gaussian perturbations. 
    
    According to \cref{eq:cmp_noise}, the function $\mathbb{\pi}^\dagger$ can be modeled by removing the degradation correlated with $\bm{y}$ and $\tilde{\bm{s}}$ from the desired image $\bm{y}$ with $\eta$. 
    {To resolve this distortion, we employ an iterative refinement strategy within the diffusion framework. Instead of attempting a one-step restoration, the model progressively decouples the target state from the physical substrate through multiple timesteps. This iterative process allows the denoiser to adaptively focus on different scales of surface-dependent residues. }
    
    Specifically, we employ IR-SDE~\cite{Luo_ICML_2023} as the backbone. Denote the desired image $\bm{y}$ as mean state $\mu$, the compensation image $\bm{y}^*$ as start state $x(0)$. Thus, in the degradation process, $y^*$ can be gradually degraded by adding noise to $\bm{y}^*$ according to the transition kernel $p_t$ and the timestep. In the $T$ timestep, the terminal state $x(T)$ arrives at the desired image $\bm{y}$ with noise $\epsilon$. 
    
    For photometric compensation SDE, as illustrated in \cref{fig:flowchart}, the compensation image $y^*$ can be recovered through iterative noise removal. The noise at the $t$-th timestep, denoted as ${\epsilon}_t$, can be predicted by the noise estimation network. Subsequently, 
    $dx_t$ can be estimated based on \cref{eq:score,eq:irsde-reverse}, and $x_{t-1}$ is then derived accordingly. 
    Through this iterative process, we perform $T$ sampling steps to obtain the final $x(0)$. {Benefiting from this iterative mechanism, our framework offers a strong capacity to capture complex nonlinear mappings through an additive trajectory, leading to a high-perceptual compensation result that matches human visual characteristics.}
    
    \subsection{Architecture of noise estimation network}
    The UNet used in \cite{Luo_ICML_2023} performs well on image restoration. Considering the computational cost of the network, they further improve the efficiency of the network by using modified nonlinear activation-free blocks, i.e., NAFBlocks~\cite{Luo_2023_CVPR}, upon which our noise estimation network is built. 
    
    Our noise estimation network takes the noisy state $x(t)$, the content condition $\mu$, the timestep $t$, and the photometry-aware condition $\tilde{\bm{s}}$ as input, and aims to learn the noise distribution introduced by the projection and capturing process. 
    
    As defined in \cref{eq:cmp_noise}, the noise distribution in the photometric compensation task is largely influenced by surface texture, reflection, ambient lighting, and projector properties. To effectively model this noise distribution, we project three reference images with varying brightness levels (white, gray and black) onto the surface under the global illumination, and capture the corresponding surface images $\{\tilde{s}_{black}, \tilde{s}_{grey}, \tilde{s}_{white}\}$. $\tilde{s}_{grey}$ has been proven to characterize environmental information by \cite{Huang_CVPR_2019}, while the combination of $\tilde{s}_{black}$ and $\tilde{s}_{white}$ is further introduced to effectively reflect the dynamic range of the projector in the current environment. In our method, these captured surface images are concatenated along the channel dimension to form a composite photometry-aware input $\tilde{\bm{s}}$. {This term implicitly incorporates the non-linear radiometric response and environmental light models, and is injected into the SDE as the physical constraint term of the projector-camera system.}
    
    Our network consists of three branches: a UNet-like main branch and two conditional branches. The main branch learns the noise distribution by integrating features from the current state and the conditions with the architecture presented by \cite{Luo_2023_CVPR} serving as the backbone. In addition to the standard encoders, decoders, middle layer, upsampling and downsampling operations, and output layer, an additional $3\times3$ convolutional layer is inserted before each encoder and decoder to ensure dimensional consistency. The conditional branches are designed to extract multi-scale content-related features and photometry-ware features, and therefore mainly consist of encoders and downsampling operations.
    Specifically, encoder/decoder modules and downsampling/upsampling modules are implemented using NAFBlocks~\cite{Luo_2023_CVPR} and shuffle operations, respectively. 
    
    At the input stage, three parallel $3\times3$ convolutional layers extract shallow features from the difference between the current state and the content condition, the content condition itself, and the photometry-aware composite image. After that, the conditional features are element-wise subtracted, and the results are injected into the UNet backbone through channel-wise concatenation with the main branch features.
    The conditional branches further extract multi-scale content-conditional and photometry-aware features via encoders and downsampling operations. The resulting features are again subtracted and injected into the backbone using the same strategy as before. To improve reconstruction quality, the subtracted features at different encoding scales are integrated into the decoder of the main branch. The details of our architecture are shown in the supplementary material. 
    
    \subsection{Compensation pipeline}
    
    During training, since ground truth compensation images are difficult to obtain, we follow the surrogate evaluation protocol presented by \cite{Huang_CVPR_2019} and employ image pairs $(\bm{y},\tilde{\bm{y}})$ instead of $(\bm{y}^*,\bm{y})$. In the forward process of the diffusion model, ${\bm{y}}$ and $\tilde{\bm{y}}$ are taken as the initial state $x(0)$ and the mean state $\mu$, respectively, while $x(t)$ is generated towards $\mu$ with additional perturbations by injecting Gaussian noise into $x(0)$ according to the cosine noise schedule. In the reverse-time SDE, the noise $\epsilon_t$ in \cref{eq:score} is predicted by the noise estimation network, after which $\hat{x}(t-1)$ is estimated through \cref{eq:irsde-reverse}, \cref{eq:score}. The network parameters are subsequently optimized by minimizing the loss defined in \cref{eq:loss-err-x}. 
    
    For evaluation on public datasets, we take $\tilde{\bm{y}}$ as the mean state $\mu$. The terminal state $x(T)$ is generated by injecting Gaussian noise into $\mu$, and the initial state $x(0)$ is then estimated through iterative reverse-time SDE. The generated compensation image is finally assessed by computing its score with respect to $\bm{y}$.

    In real-world applications, ${\bm{y}}$ is adopted as the mean state $\mu$. While the other procedures remain consistent with the evaluation stage. The resulting initial state $x(0)$ is taken as the compensation image to be projected. This image is then rendered onto the current environment to obtain the actual projection outcome. Accordingly, in the real-projection experiments, we compare the resulting display effects with the desired image $\bm{y}$ to further evaluate the performance of the algorithm.
    

    \section{Experiments}
    \subsection{Datasets}
    {Our experiments are conducted on both public datasets and real-world projections to provide a comprehensive evaluation.} 

    {\textbf{Public datasets:} We utilize two widely-used real photometric compensation datasets\cite{Huang_CVPR_2019,Huang_PAMI_2021} and a synthetic dataset\cite{Huang_PAMI_2021} to train and evaluate models. The combined three datasets consist of 143 setups, each containing 500 training pairs and 200 testing pairs. We utilized the training pairs from 130 setups with different environment configurations to pre-train our model, and the remaining 13 setups (7 from the CompenNet dataset\cite{Huang_CVPR_2019} and 6 from CompenNeSt\cite{Huang_PAMI_2021}) are reserved as the testing set. For each setup in the testing set, the training image pairs are utilized to fine-tune models, while the corresponding testing image pairs are used for performance evaluation. Notably, this testing set involves environmental configurations that were entirely excluded from the training sets, thereby ensuring an assessment of the per-trained models' generalization to unseen scenarios.}
    
    {\textbf{Real-world projection:} To verify the effectiveness of our algorithm in real-world scenarios, we projected the compensated images onto actual surfaces and evaluated the resulting visual performance. To simulate the photometric complexity of real-world scenarios while decoupling the impact of geometric distortions, we printed various natural textures (e.g., leaf, wall, cardboard, and painting patterns) onto eight flat PVC panels.} These textured surfaces were positioned $0.8-1.2$ meters from the projector, with cameras placed within a $0.5$ meter radius around the projector. We then projected 500 ideal training images of the public dataset\cite{Huang_CVPR_2019} onto these surfaces using an EPSON CB-X05 projector and captured the corresponding displayed results using a Sony $\alpha 6400$ camera to construct training image pairs. Furthermore, to simulate the specular surfaces, we also placed a piece of the transparent plastic sheet in front of two of these projection surfaces. {Additionally, to implement geometric correction, we projected a chessboard pattern onto each surface and established the mapping by computing the homography matrix between the projected and the displayed image pair. After geometric correction, all captured images were corrected to a frontal view perspective with the size of $256\times256$. }

    {For a fair comparison, we used the same training and testing data for all methods. To improve robustness, the training data were augmented with random flips and rotations.}

    \subsection{Evaluation metrics}
    We evaluate our method on three common image quality metrics, PSNR, SSIM \cite{Zhou_TIP_2004}, and {CIE standard for perceptual color differences($\Delta E$)\cite{Sharma2005TheCC}}. Additionally, we also use LIPIS \cite{Zhang_CVPR_2018} and Fréchet Inception Distance (FID) \cite{Heusel_2017_nips} to assess the visual perceptual performance of compensation effects.

    \subsection{Implementation details}
    We employ the cosine noise schedule introduced in \cite{nichol_imcl_2021}:
    \begin{equation}
        \theta_t = 1-\frac{f(t)}{f(0)}, ~ f(x)=\cos\left(\frac{t/T+s}{1+s}\cdot\frac{\pi}{2}\right)^2, 
    \end{equation}
    where $s=0.008$ following \cite{nichol_imcl_2021}. The number of steps $T$ is set to 100 in all experiments. 
    
    Our method is implemented in PyTorch and executed on an NVIDIA GeForce RTX 4090 GPU. During pre-training, we employ the Lion optimizer\cite{Chen_nips_2023} with $\beta_1 = 0.9 $ and $ \beta_2 = 0.99$ to train the model for 200,000 iterations. The initial learning rate is set to 4e-5 and decayed by a factor of 0.1 every 100,000 iterations. The batch size is set to 6. 
    For CNN-based methods, the initial learning rate is also set to 1e-3, with training performed for 100,000 iterations. The learning rate is decayed by a factor of 0.2 after 60,000 iterations, and the batch size is set to 64. 
    For fine-tuning, we adopt a learning rate of 1e-5 with 1,000 iterations and a batch size of 6. For comparison, CNN-based methods are finetuned for 800 iterations with a learning rate of 1e-3 and a batch size of 64\footnote{{To balance the increase in training data, we scale the batch size from the original 8 to 64.}}.

    \subsection{{Comparison with the state-of-the-art methods}}
     We evaluated our DiffPC against three state-of-the-art CNN-based photometric compensation approaches: CompenNet\cite{Huang_CVPR_2019}, CompenNeSt\cite{Huang_PAMI_2021}, and PANet\cite{Wang_VR_2023}. {To measure the algorithms' generalization capability in unknown environments and their adaptation efficiency through fine-tuning, we assessed each method's performance using both pre-trained and fine-tuned models in both public datasets and real-world evaluations. 
     Specifically, all pre-trained models are directly applied to the 200 testing image pairs of the testing set or real scenarios with unknown surfaces\footnote{The corresponding results are denoted with the ``-pre'' suffix.}. We fine-tune the pre-trained models using the 500 training image pairs of each current setup. The fine-tuned models are then evaluated on the 200 testing image pairs of the current setup\footnote{The corresponding results are denoted with the ``-ft'' suffix.}.}
    
    \subsubsection{Experiments on public datasets}
    \label{sec:pubexp}
    All methods are evaluated on 13 setups selected from two public datasets\cite{Huang_CVPR_2019,Huang_PAMI_2021}.
    
    From the quantitative results of pre-trained and fine-tuned models presented in \cref{tab:pub_pre} and \cref{tab:pub_ft}, our DiffPC achieves notable improvements in perceptual quality. In \cref{tab:pub_pre}, our method outperforms all CNN-based approaches across all evaluation metrics. In particular, PANet, designed for high-resolution compensation, delivers suboptimal performance on low-resolution inputs due to its shuffle-based acceleration mechanism. {These results initially demonstrate the superior generalization capability of our method in the unseen scenarios.}
    During fine-tuning, as CompenNeSt failed to converge on one setup, we report the average results excluding it in \cref{tab:pub_ft}. DiffPC maintains a significant performance advantage after fine-tuning with a few iterations. Notably, our method, whether in its pre-trained or fine-tuned form, shows a substantial advantage over CNN-based approaches in terms of FID scores, highlighting its superior capability in modeling perceptual quality. These results further demonstrate the adaptation efficiency and superior compensation quality of our method.‌  
    \begin{table*}[!htb]
        \centering
        \caption{Results of pre-trained models on two public datasets. (The best results are highlighted in red.)}
        \label{tab:pub_pre}
        \small
        \setlength{\tabcolsep}{3pt}
        \begin{tabular}{l|ccccc|ccccc|ccccc}
         \hline
         & \multicolumn{5}{c|}{CompenNet Datasets} & \multicolumn{5}{c|}{CompenNeSt Datasets} &\multicolumn{5}{c}{Average} \\
         \cline{2-16}
         & PSNR $\uparrow$ & SSIM & FID $\downarrow$ & LPIPS $\downarrow$ & $\Delta E$ $\downarrow$ &  PSNR $\uparrow$ & SSIM $\uparrow$ & FID $\downarrow$ & LPIPS $\downarrow$ & $\Delta E$ $\downarrow$ & PSNR $\uparrow$ & SSIM $\uparrow$ & FID $\downarrow$ & LPIPS $\downarrow$ & $\Delta E$ $\downarrow$ \\
        \hline
           Uncompensated & 12.2583	& 0.5078 &	245.4870 &	0.4378 &	22.8074 & 12.2970	& 0.5100 &	239.8782 &	0.4426 &	22.1011 & 12.2762 &	0.5088	& 242.8983	& 0.4400	& 22.4814 \\
           CompenNet-pre & 15.5382 & 0.6530 & 126.0144 & 0.3531 & 14.7136 & 18.9877 & 0.6905 & 122.6027 & 0.2944 & 10.8374 & 17.1303 & 0.6703 & 124.4398 & 0.3261 & 12.9246\\
           CompenNeSt-pre & 15.9188 & 0.6659 & 114.2655 & 0.3417 & 14.4087 & 18.9668 & 0.7015 & 117.0248 & 0.2872 & 10.6231 & 17.3256 & 0.6823 & 115.5390 & 0.3165 & 12.6615\\ 
           PANet-pre & 14.0817 & 0.5916 & 127.8356 & 0.3887 & 16.9488 & 16.7775 & 0.6373 & 129.7054 & 0.3441 & 13.1814 & 15.3259 & 0.6127 & 128.6986 & 0.3681 & 15.2100 \\ 
           DiffPC-pre & \textcolor{red}{\textbf{17.1614}} & \textcolor{red}{\textbf{0.6920}} & \textcolor{red}{\textbf{78.7853}} & \textcolor{red}{\textbf{0.3027}} & \textcolor{red}{\textbf{12.2787}} & \textcolor{red}{\textbf{21.0479}} & \textcolor{red}{\textbf{0.7084}}	& \textcolor{red}{\textbf{72.7515}} & \textcolor{red}{\textbf{0.2116}} & \textcolor{red}{\textbf{8.0062}}  & \textcolor{red}{\textbf{18.9552}} & \textcolor{red}{\textbf{0.6996}} & \textcolor{red}{\textbf{76.0005}} & \textcolor{red}{\textbf{0.2606}} & \textcolor{red}{\textbf{10.3067}} \\
      \hline
      \end{tabular}
    \end{table*}

    \begin{table*}[!htb]
        \centering
        \caption{{Results of fine-tuned models on two public datasets\protect\footnotemark. (The best results are highlighted in red.)}}
        \label{tab:pub_ft}
        \small
        \setlength{\tabcolsep}{3pt}
        
        \begin{tabular}{l|ccccc|ccccc|ccccc}
        \hline
         & \multicolumn{5}{c|}{CompenNet Datasets} & \multicolumn{5}{c|}{CompenNeSt Datasets} &\multicolumn{5}{c}{Average} \\
         \cline{2-16}
         & PSNR $\uparrow$ & SSIM & FID $\downarrow$ & LPIPS $\downarrow$ & $\Delta E$ $\downarrow$ &  PSNR $\uparrow$ & SSIM $\uparrow$ & FID $\downarrow$ & LPIPS $\downarrow$ & $\Delta E$ $\downarrow$ & PSNR $\uparrow$ & SSIM $\uparrow$ & FID $\downarrow$ & LPIPS $\downarrow$ & $\Delta E$ $\downarrow$ \\
         \hline
           Uncompensated & 12.4891	& 0.5290 &	235.7780 &	0.4291	& 21.9133 & 12.2970	& 0.5100 &	239.8782 &	0.4426 &	22.1011 & 12.3930 &	0.5195 &	237.8281 &	0.4358	& 22.0072\\
           CompenNeSt-ft & 21.9815	& \textcolor{red}{\textbf{0.7852}} &	84.1933	& 0.2061 &	7.8857 & 21.5593 &	\textcolor{red}{\textbf{0.7469}}	 & 92.8450 &	0.2176 &	8.3436 & 21.7704 &	\textcolor{red}{\textbf{0.7660}} &	88.5191 &	0.2119 &	8.1147 \\ 
           PANet-ft & 21.4374	& 0.7488 &	99.3065 &	0.2290 &	8.4881 & 21.2724	& 0.7150 &	104.0446 &	0.2359 &	8.6795 & 21.3549 &	0.7319 &	101.6756 &	0.2325 &	8.5838\\ 
           DiffPC-ft & \textcolor{red}{\textbf{22.6386}}	& {0.7710}	& \textcolor{red}{\textbf{61.6590}}	& \textcolor{red}{\textbf{0.1719}} & \textcolor{red}{\textbf{6.7900}}	& \textcolor{red}{\textbf{22.2912}}	& {0.7446}	& \textcolor{red}{\textbf{66.3860}} & \textcolor{red}{\textbf{0.1880}}	& \textcolor{red}{\textbf{7.2084}}	& \textcolor{red}{\textbf{22.4649}}	&  {0.7578} &  \textcolor{red}{\textbf{64.0225}} &  \textcolor{red}{\textbf{0.1799}} &  \textcolor{red}{\textbf{6.9992}}\\
      \hline
      \end{tabular}
    \end{table*}
    \footnotetext{\scriptsize{{The results for CompenNet are excluded from \cref{tab:pub_ft} as it failed to yield valid outputs on more than half of the test setups. Additionally, due to the failure of CompenNeSt on one setup, we report the performance of all compared methods based on the remaining 12 successful setups in \cref{tab:pub_ft}. }}}

    \begin{figure*}[!htb]
      \centering
      \tiny
      
      \begin{minipage}[t]{0.097\textwidth}
        \centering
        {Surface}
      \end{minipage}
      \begin{minipage}[t]{0.097\textwidth}
        \centering
        {CompenNet-pre}
      \end{minipage}
      \begin{minipage}[t]{0.097\textwidth}
        \centering
        {CompenNeSt-pre}
      \end{minipage}
      \begin{minipage}[t]{0.097\textwidth}
        \centering
        {PANet-pre}
      \end{minipage}
      \begin{minipage}[t]{0.097\textwidth}
        \centering
        {DiffPC-pre}
      \end{minipage}
      \begin{minipage}[t]{0.097\textwidth}
        \centering
        {CompenNet-ft}
      \end{minipage}
      \begin{minipage}[t]{0.097\textwidth}
        \centering
        {CompenNeSt-ft}
      \end{minipage}
      \begin{minipage}[t]{0.097\textwidth}
        \centering
        {PANet-ft}
      \end{minipage}
      \begin{minipage}[t]{0.097\textwidth}
        \centering
        {DiffPC-ft}
      \end{minipage}
      \begin{minipage}[t]{0.097\textwidth}
        \centering
        {GT}
      \end{minipage}
      
      \vskip 0.2cm
      \begin{minipage}[t]{0.097\textwidth}
        \centering
        \includegraphics[width=\linewidth]{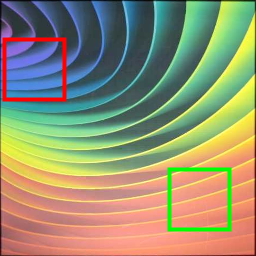}\\
        \includegraphics[width=\linewidth]{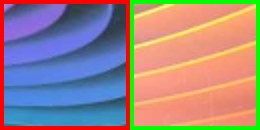}\\
        \includegraphics[width=\linewidth]{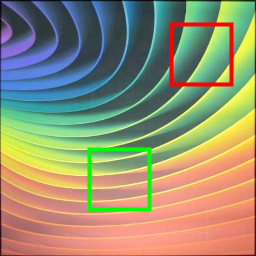}\\
        \includegraphics[width=\linewidth]{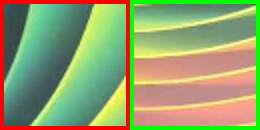}\\
      \end{minipage}
      \begin{minipage}[t]{0.097\textwidth}
        \centering
        \includegraphics[width=\linewidth]{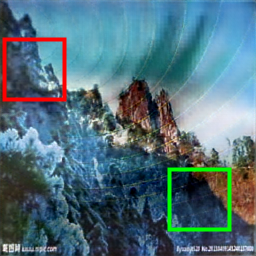}\\
        \includegraphics[width=\linewidth]{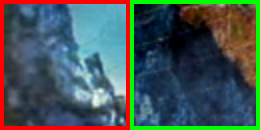}\\
        \includegraphics[width=\linewidth]{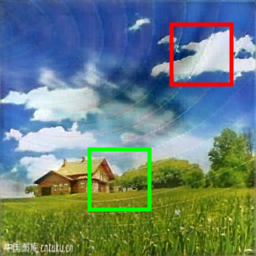}\\
        \includegraphics[width=\linewidth]{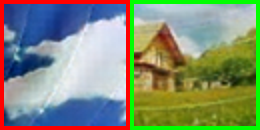}
      \end{minipage}
      \begin{minipage}[t]{0.097\textwidth}
        \centering
        \includegraphics[width=\linewidth]{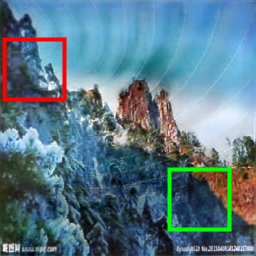}
        \includegraphics[width=\linewidth]{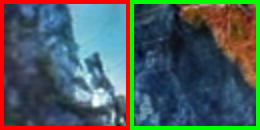}
        \includegraphics[width=\linewidth]{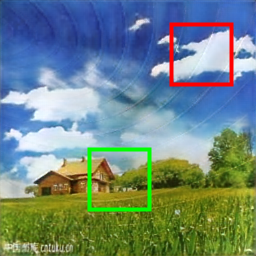}
        \includegraphics[width=\linewidth]{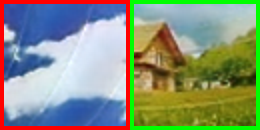}
      \end{minipage}
      \begin{minipage}[t]{0.097\textwidth}
        \centering
        \includegraphics[width=\linewidth]{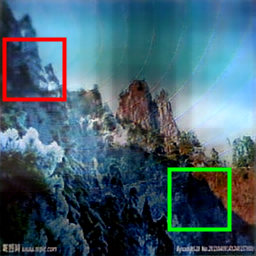}
        \includegraphics[width=\linewidth]{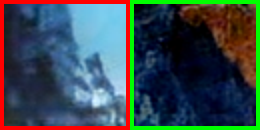}
        \includegraphics[width=\linewidth]{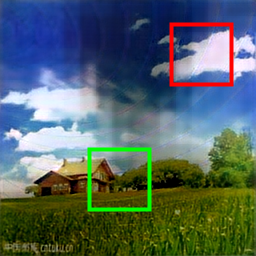}
        \includegraphics[width=\linewidth]{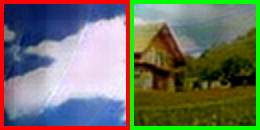}
      \end{minipage}
      \begin{minipage}[t]{0.097\textwidth}
        \centering
        \includegraphics[width=\linewidth]{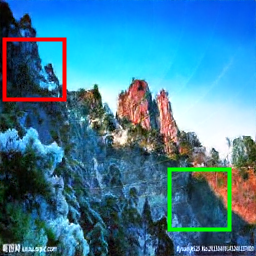}
        \includegraphics[width=\linewidth]{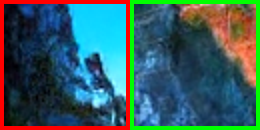}
        \includegraphics[width=\linewidth]{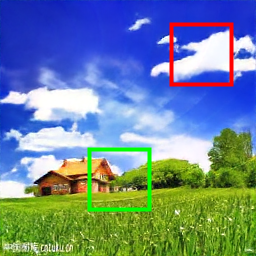}
        \includegraphics[width=\linewidth]{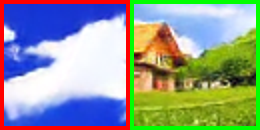}
      \end{minipage}
      \begin{minipage}[t]{0.097\textwidth}
        \centering
        \includegraphics[width=\linewidth]{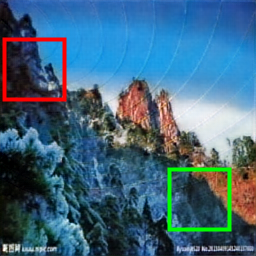}
        \includegraphics[width=\linewidth]{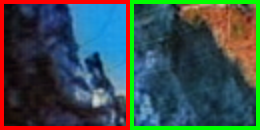}
        \includegraphics[width=\linewidth]{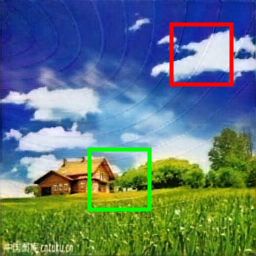}
        \includegraphics[width=\linewidth]{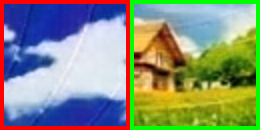}
      \end{minipage}
      \begin{minipage}[t]{0.097\textwidth}
        \centering
        \includegraphics[width=\linewidth]{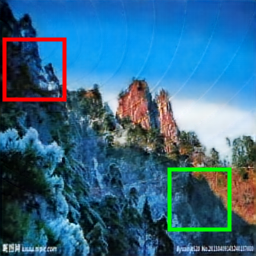}
        \includegraphics[width=\linewidth]{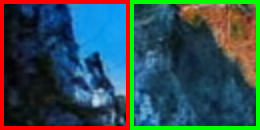}
        \includegraphics[width=\linewidth]{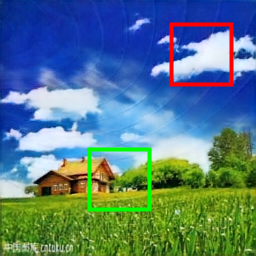}
        \includegraphics[width=\linewidth]{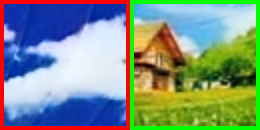} 
      \end{minipage}
      \begin{minipage}[t]{0.097\textwidth}
        \centering
        \includegraphics[width=\linewidth]{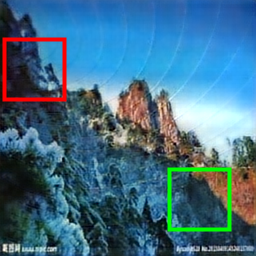}
        \includegraphics[width=\linewidth]{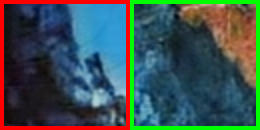}
        \includegraphics[width=\linewidth]{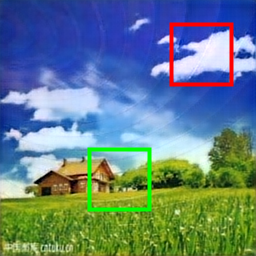}
        \includegraphics[width=\linewidth]{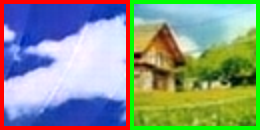}
      \end{minipage}
      \begin{minipage}[t]{0.097\textwidth}
        \centering
        \includegraphics[width=\linewidth]{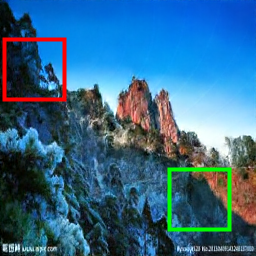}
        \includegraphics[width=\linewidth]{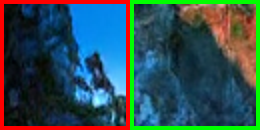}
        \includegraphics[width=\linewidth]{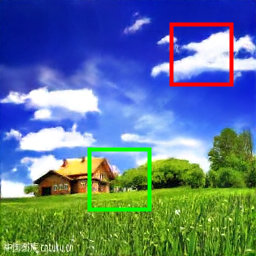}
        \includegraphics[width=\linewidth]{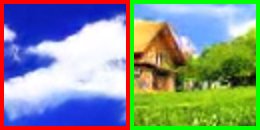}
      \end{minipage}
      \begin{minipage}[t]{0.097\textwidth}
        \centering
        \includegraphics[width=\linewidth]{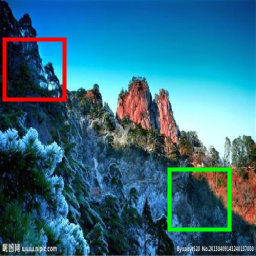}
        \includegraphics[width=\linewidth]{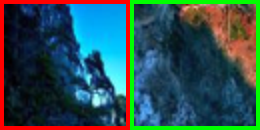}
        \includegraphics[width=\linewidth]{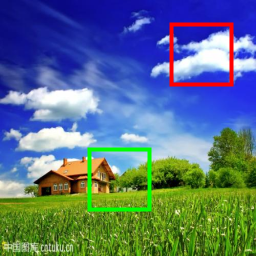}
        \includegraphics[width=\linewidth]{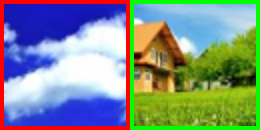}
      \end{minipage}

      \vskip 0.2cm
      \begin{minipage}[t]{0.097\textwidth}
        \centering
        \includegraphics[width=\linewidth]{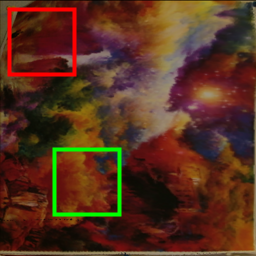}\\
        \includegraphics[width=\linewidth]{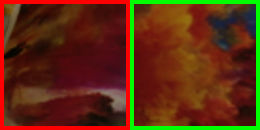}\\
        \includegraphics[width=\linewidth]{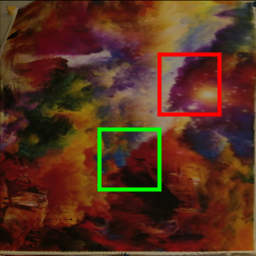}\\
        \includegraphics[width=\linewidth]{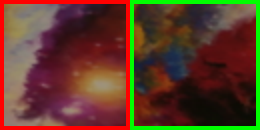}\\
      \end{minipage}
      \begin{minipage}[t]{0.097\textwidth}
        \centering
        \includegraphics[width=\linewidth]{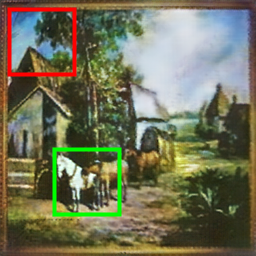}
        \includegraphics[width=\linewidth]{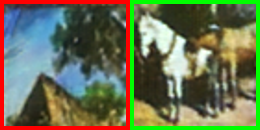}
        \includegraphics[width=\linewidth]{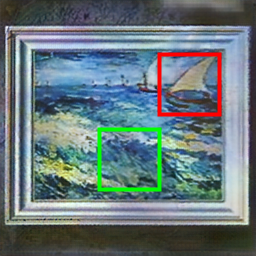}
        \includegraphics[width=\linewidth]{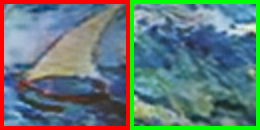}
      \end{minipage}
      \begin{minipage}[t]{0.097\textwidth}
        \centering
        \includegraphics[width=\linewidth]{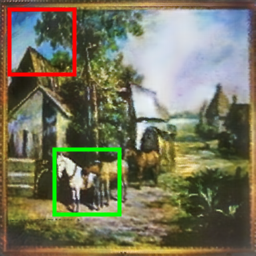}
        \includegraphics[width=\linewidth]{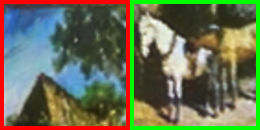}
        \includegraphics[width=\linewidth]{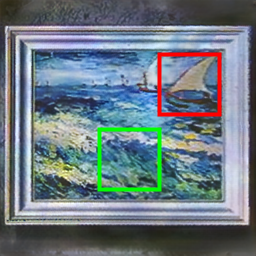}
        \includegraphics[width=\linewidth]{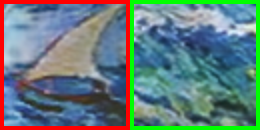}
      \end{minipage}
      \begin{minipage}[t]{0.097\textwidth}
        \centering
        \includegraphics[width=\linewidth]{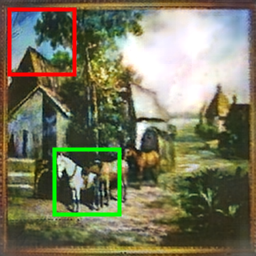}
        \includegraphics[width=\linewidth]{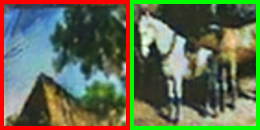}
        \includegraphics[width=\linewidth]{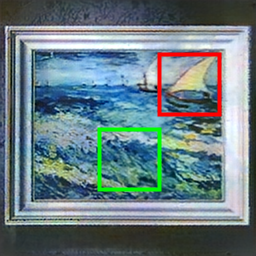}
        \includegraphics[width=\linewidth]{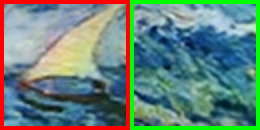}
      \end{minipage}
      \begin{minipage}[t]{0.097\textwidth}
        \centering
        \includegraphics[width=\linewidth]{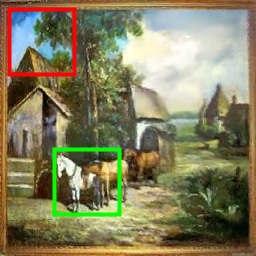}
        \includegraphics[width=\linewidth]{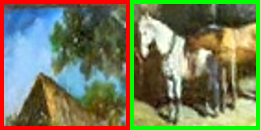}
        \includegraphics[width=\linewidth]{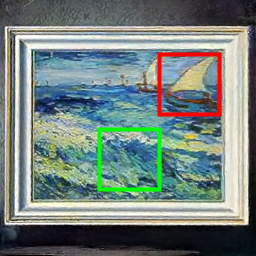}
        \includegraphics[width=\linewidth]{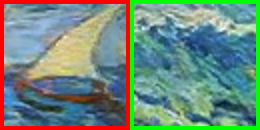}
      \end{minipage}
      \begin{minipage}[t]{0.097\textwidth}
        \centering
        \includegraphics[width=\linewidth]{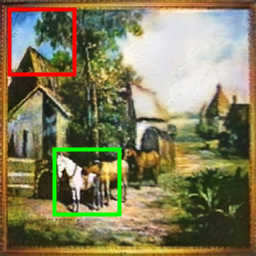}
        \includegraphics[width=\linewidth]{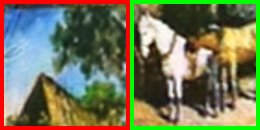}
        \includegraphics[width=\linewidth]{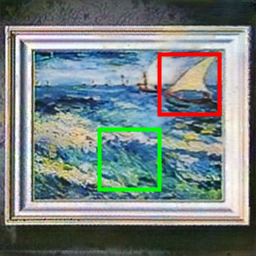}
        \includegraphics[width=\linewidth]{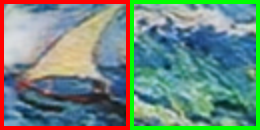}
      \end{minipage}
      \begin{minipage}[t]{0.097\textwidth}
        \centering
        \includegraphics[width=\linewidth]{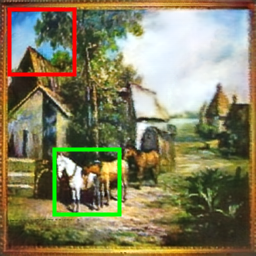}
        \includegraphics[width=\linewidth]{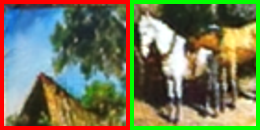}
        \includegraphics[width=\linewidth]{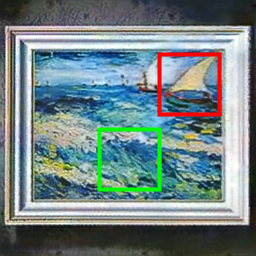}
        \includegraphics[width=\linewidth]{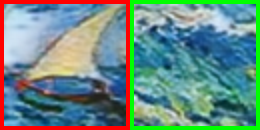}
      \end{minipage}
      \begin{minipage}[t]{0.097\textwidth}
        \centering
        \includegraphics[width=\linewidth]{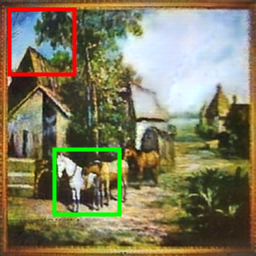}
        \includegraphics[width=\linewidth]{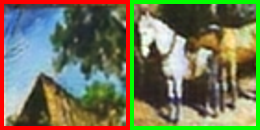}
        \includegraphics[width=\linewidth]{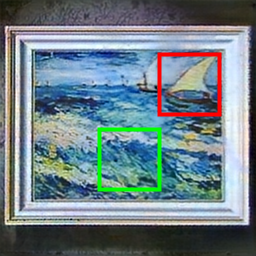}
        \includegraphics[width=\linewidth]{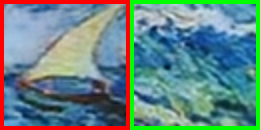}
      \end{minipage}
      \begin{minipage}[t]{0.097\textwidth}
        \centering
        \includegraphics[width=\linewidth]{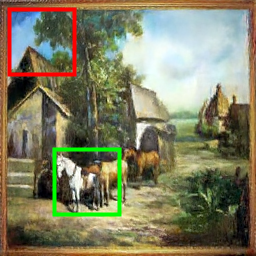}
        \includegraphics[width=\linewidth]{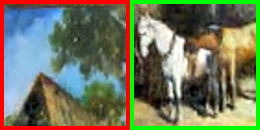}
        \includegraphics[width=\linewidth]{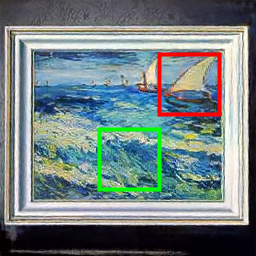}
        \includegraphics[width=\linewidth]{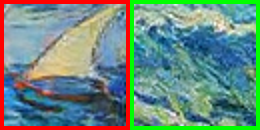}
      \end{minipage}
      \begin{minipage}[t]{0.097\textwidth}
        \centering
        \includegraphics[width=\linewidth]{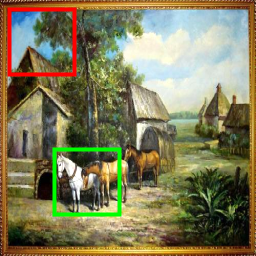}
        \includegraphics[width=\linewidth]{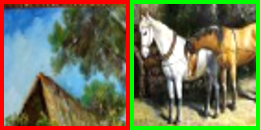}
        \includegraphics[width=\linewidth]{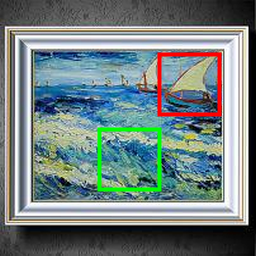}
        \includegraphics[width=\linewidth]{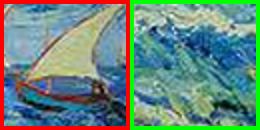}
      \end{minipage}
  
      \caption{Qualitative compensation results on the public datasets. Top two rows: CompenNet dataset~\cite{Huang_CVPR_2019}; bottom two rows: CompenNeSt dataset~\cite{Huang_PAMI_2021}. Columns (left to right): the captured surface, compensation results from pre-trained CompenNet, CompenNeSt, and PANet, our pre-trained model, fine-tuned CompenNet, CompenNeSt, and PANet, our fine-tuned model, and finally the desired ground-truth images.}
      \label{fig:compennet}
    \end{figure*}

     Additionally, \cref{fig:compennet} presents the comparison of qualitative compensation results. Under the surrogate evaluation protocol, the final outputs of all methods converge toward the desired image $\bm{y}$. As illustrated in \cref{fig:compennet}, the CNN-based methods generate compensation images with noticeable brightness inconsistencies, leading to degraded perceptual quality. Our method produces a more uniform brightness distribution, resulting in good visual perception. This advantage is particularly evident in the compensation results obtained by the pre-trained models.
     This is attributed to the fact that CNN-based methods typically predict compensation images in a single step, with optimization often guided by pixel-wise losses that fail to explicitly account for human perceptual characteristics. In contrast, our diffusion-based method restores images through an iterative denoising process. This gradual refinement resembles the way that the human visual system incrementally constructs perceptual representations, thereby preserving both local coherence and global consistency. Moreover, distortions in the projection process arise from complex nonlinear mechanisms, which are difficult for simple CNN-based methods to model. By learning the distributions of high-dimensional data, diffusion models produce results that better reflect the statistical properties of the data, thus yielding superior generalization performance and . 
     ‌

    \subsubsection{Real-World Projection Experiments}
    The previous subsection evaluated the performance of the algorithms on public datasets using the surrogate evaluation protocol. In this subsection, we further assess the practical effectiveness of DiffPC and other CNN-based methods by conducting real-world projection experiments and comparing display results of each method in real-world scenarios. Specifically, 
    compensation images were generated using both the pre-trained and fine-tuned models and projected onto the corresponding target surface to produce the final visual effects. 
    
    As reported in \cref{tab:real}, 
    both our pre-trained and fine-tuned models outperforms CNN-based approaches in real projections. In particular, our pre-trained models even achieves lower FID score than all fine-tuned CNN-based methods. These results further demonstrate the strong generalization capability and practical effectiveness of our method.

     \begin{table}[!htb]
        \centering
        \small
        \caption{The average results on real-world projection. (The best results of the pre-trained model are highlighted in blue, and the best results after fine-tuning are highlighted in red.)}
        \label{tab:real}
        \begin{tabular}{lccccc}
         \toprule
         & PSNR $\uparrow$ & SSIM $\uparrow$ & FID $\downarrow$ & LPIPS $\downarrow$ & $\Delta E$ $\downarrow$\\
        \midrule
         Uncompensated  &  12.4899	& 0.5147	& 191.1356	& 0.4249 & 21.9904\\
         \hline
           CompenNet-pre  &  17.3495	& 0.6446	& 128.0260	& 0.3052	& 12.4729\\ 
           CompenNeSt-pre  &  17.0110	& 0.6319	& 135.0019	& 0.3088  & 12.9338 \\
           PANet-pre  &  18.0439	& 0.6249	& 138.6146	& 0.2965	& 12.2256\\
           DiffPC-pre  &  \textcolor{blue}{\textbf{20.5011}}	& \textcolor{blue}{\textbf{0.6843}}	& \textcolor{blue}{\textbf{109.1843}}	& \textcolor{blue}{\textbf{0.2590}}	& \textcolor{blue}{\textbf{9.7514}}\\
           \hline
           CompenNet-ft  &  21.4305	& 0.6804	& 114.4265	& 0.2163	& 8.9021\\ 
           CompenNeSt-ft  & 21.3460	& 0.6851	& 110.2675	& \textcolor{red}{\textbf{0.2111}}  & 8.6365 \\
           PANet-ft  &  21.3610	& 0.6570	& 123.5151	& 0.2196 & 8.8263	\\
           DiffPC-ft  &  \textcolor{red}{\textbf{22.0148}}	 & \textcolor{red}{\textbf{0.7096}}	& \textcolor{red}{\textbf{102.9548}}	& 0.2136 & \textcolor{red}{\textbf{8.2338}}		\\
      \bottomrule
      \end{tabular}
    \end{table}
    The qualitative comparisons in \cref{fig:realprojection} indicate that the display effects produced by pre-trained models exhibit noticeable discrepancies in color and brightness compared to the ideal images. After fine-tuning, all methods get significant improvements in display quality. However, our approach maintains better performance than CNN-based approaches by effectively suppressing background texture interference and enhancing visual perception performance.‌
    In addition, the last two rows of \cref{fig:realprojection} illustrate the performance under specular conditions. Pre-trained models of CompenNet and CompenNeSt are strongly affected by specular highlights, leading to color distortions and local brightness inconsistencies. In contrast, our method and PANet exhibit greater robustness to such conditions, with our approach achieving superior visual consistency and overall display quality compared to PANet. These results further validate the adaptability and perceptual advantages of our diffusion-based photometric compensation framework under diverse real-world scenarios.

\begin{figure*}[!htbp]
  \centering
  
  \tiny
  
  \begin{minipage}[t]{0.088\textwidth}
    \centering
    {Surface}
  \end{minipage}
  \begin{minipage}[t]{0.088\textwidth}
    \centering
    {Uncompensated}
  \end{minipage}
  \begin{minipage}[t]{0.088\textwidth}
    \centering
    {CompenNet-pre}
  \end{minipage}
  \begin{minipage}[t]{0.088\textwidth}
    \centering
    {CompenNeSt-pre}
  \end{minipage}
  \begin{minipage}[t]{0.088\textwidth}
    \centering
    {PANet-pre}
  \end{minipage}
  \begin{minipage}[t]{0.088\textwidth}
    \centering
    {DiffPC-pre}
  \end{minipage}
  \begin{minipage}[t]{0.088\textwidth}
    \centering
    {CompenNet-ft}
  \end{minipage}
  \begin{minipage}[t]{0.088\textwidth}
    \centering
    {CompenNeSt-ft}
  \end{minipage}
  \begin{minipage}[t]{0.088\textwidth}
    \centering
    {PANet-ft}
  \end{minipage}
  \begin{minipage}[t]{0.088\textwidth}
    \centering
    {DiffPC-ft}
  \end{minipage}
  \begin{minipage}[t]{0.088\textwidth}
    \centering
    {GT}
  \end{minipage}
  \vskip 0.2cm
  \begin{minipage}[t]{0.088\textwidth}
    \centering
    \includegraphics[width=\linewidth]{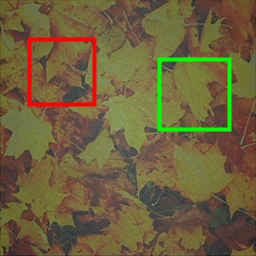}\\
    \includegraphics[width=\linewidth]{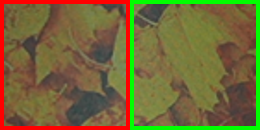}\\
    \includegraphics[width=\linewidth]{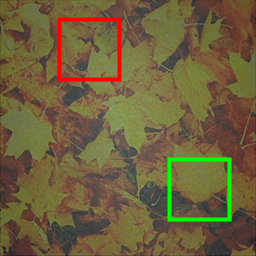}\\
    \includegraphics[width=\linewidth]{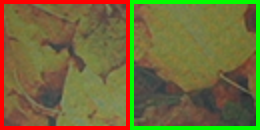}\\
  \end{minipage}
  \begin{minipage}[t]{0.088\textwidth}
    \centering
    \includegraphics[width=\linewidth]{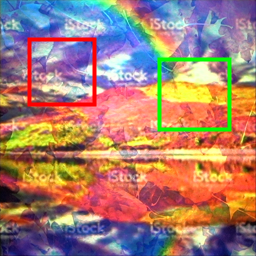}\\
    \includegraphics[width=\linewidth]{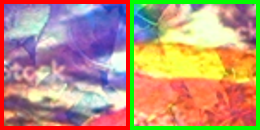}\\
    \includegraphics[width=\linewidth]{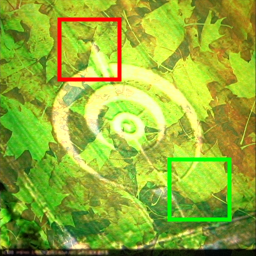}\\
    \includegraphics[width=\linewidth]{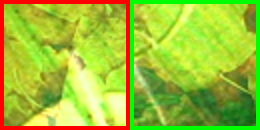}\\
  \end{minipage}
  \begin{minipage}[t]{0.088\textwidth}
    \centering
    \includegraphics[width=\linewidth]{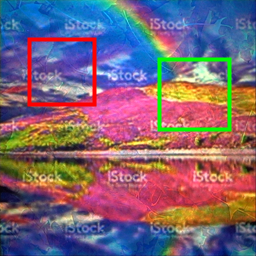}
    \includegraphics[width=\linewidth]{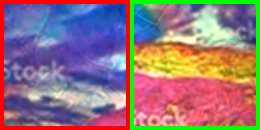}
    \includegraphics[width=\linewidth]{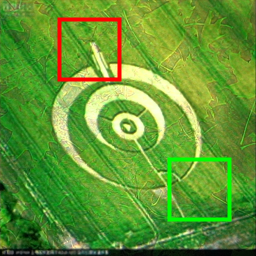}
    \includegraphics[width=\linewidth]{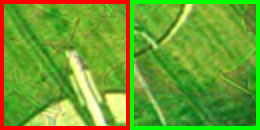}
  \end{minipage}
  \begin{minipage}[t]{0.088\textwidth}
    \centering
    \includegraphics[width=\linewidth]{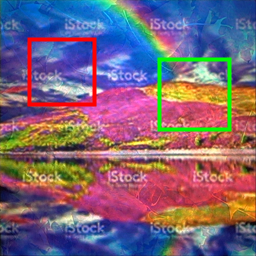}
    \includegraphics[width=\linewidth]{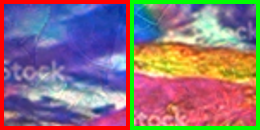}
    \includegraphics[width=\linewidth]{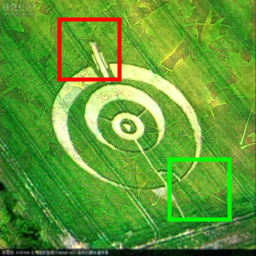}
    \includegraphics[width=\linewidth]{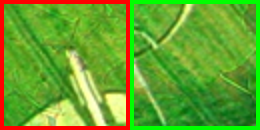}
  \end{minipage}
  \begin{minipage}[t]{0.088\textwidth}
    \centering
    \includegraphics[width=\linewidth]{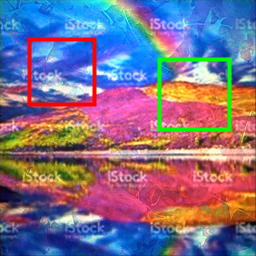}
    \includegraphics[width=\linewidth]{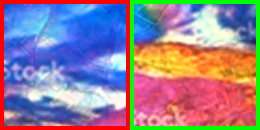}
    \includegraphics[width=\linewidth]{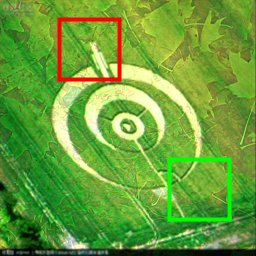}
    \includegraphics[width=\linewidth]{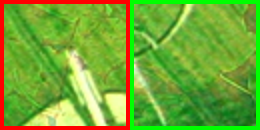}
  \end{minipage}
  \begin{minipage}[t]{0.088\textwidth}
    \centering
    \includegraphics[width=\linewidth]{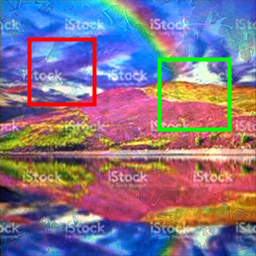}
    \includegraphics[width=\linewidth]{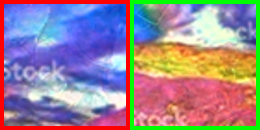}
    \includegraphics[width=\linewidth]{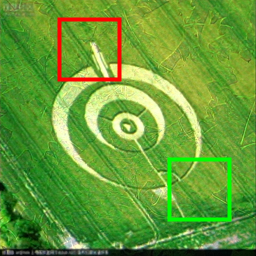}
    \includegraphics[width=\linewidth]{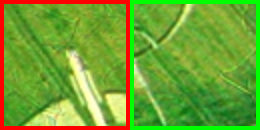}
  \end{minipage}
  \begin{minipage}[t]{0.088\textwidth}
    \centering
    \includegraphics[width=\linewidth]{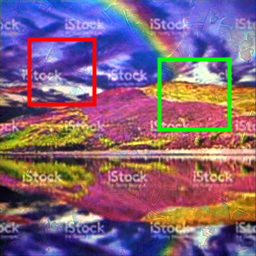}
    \includegraphics[width=\linewidth]{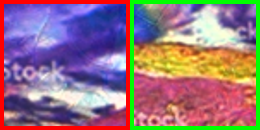}
    \includegraphics[width=\linewidth]{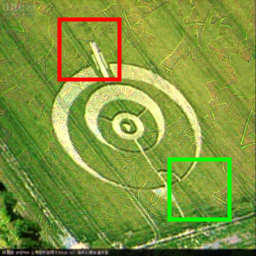}
    \includegraphics[width=\linewidth]{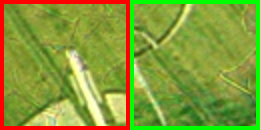}
  \end{minipage}
  \begin{minipage}[t]{0.088\textwidth}
    \centering
    \includegraphics[width=\linewidth]{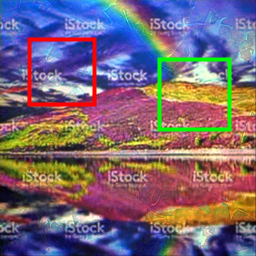}
    \includegraphics[width=\linewidth]{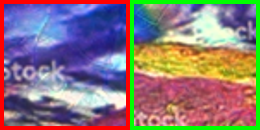}
    \includegraphics[width=\linewidth]{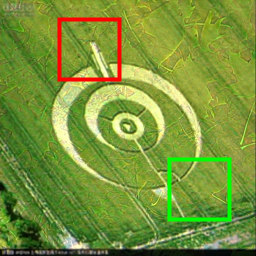}
    \includegraphics[width=\linewidth]{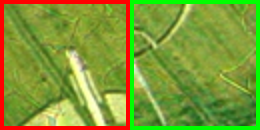}
  \end{minipage}
  \begin{minipage}[t]{0.088\textwidth}
    \centering
    \includegraphics[width=\linewidth]{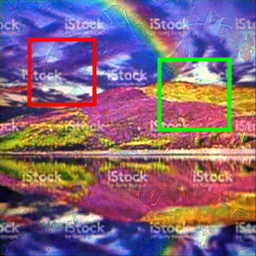}
    \includegraphics[width=\linewidth]{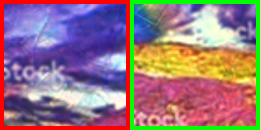}
    \includegraphics[width=\linewidth]{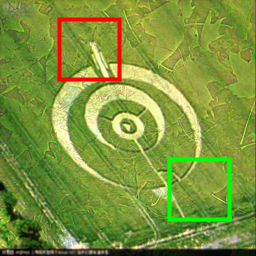}
    \includegraphics[width=\linewidth]{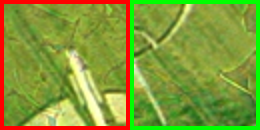}
  \end{minipage}
  \begin{minipage}[t]{0.088\textwidth}
    \centering
    \includegraphics[width=\linewidth]{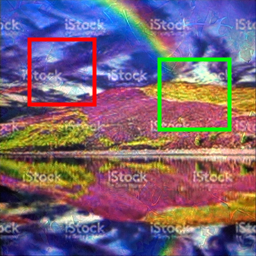}
    \includegraphics[width=\linewidth]{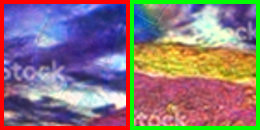}
    \includegraphics[width=\linewidth]{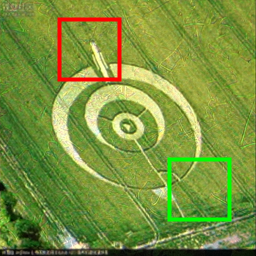}
    \includegraphics[width=\linewidth]{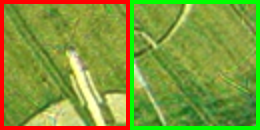}
  \end{minipage}
  \begin{minipage}[t]{0.088\textwidth}
    \centering
    \includegraphics[width=\linewidth]{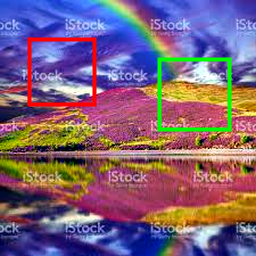}
    \includegraphics[width=\linewidth]{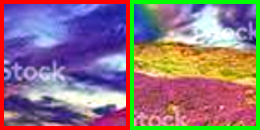}
    \includegraphics[width=\linewidth]{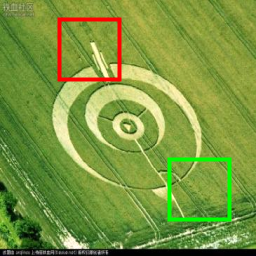}
    \includegraphics[width=\linewidth]{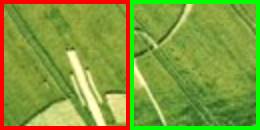}
    
  \end{minipage}
  \vskip 0.2cm
  \begin{minipage}[t]{0.088\textwidth}
    \centering
    \includegraphics[width=\linewidth]{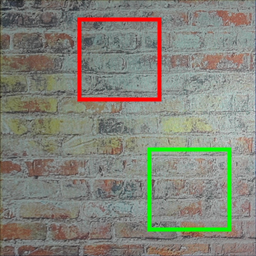}\\
    \includegraphics[width=\linewidth]{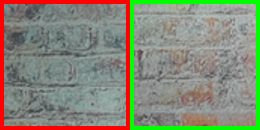}\\
    \includegraphics[width=\linewidth]{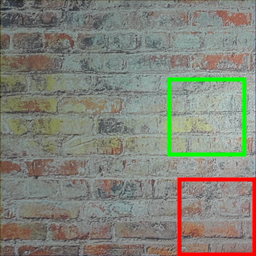}\\
    \includegraphics[width=\linewidth]{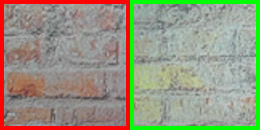}\\
  \end{minipage}
  \begin{minipage}[t]{0.088\textwidth}
    \centering
    \includegraphics[width=\linewidth]{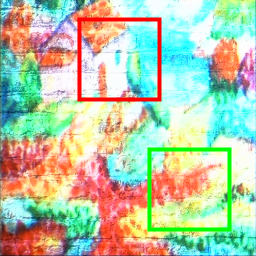}\\
    \includegraphics[width=\linewidth]{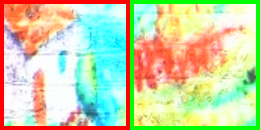}\\
    \includegraphics[width=\linewidth]{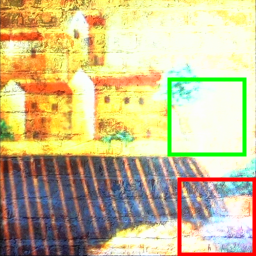}\\
    \includegraphics[width=\linewidth]{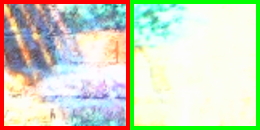}\\
  \end{minipage}
  \begin{minipage}[t]{0.088\textwidth}
    \centering
    \includegraphics[width=\linewidth]{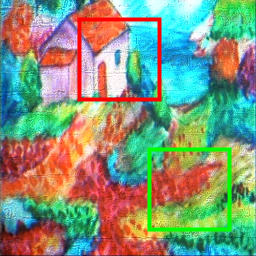}
    \includegraphics[width=\linewidth]{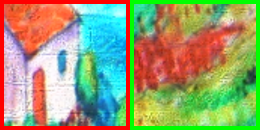}
    \includegraphics[width=\linewidth]{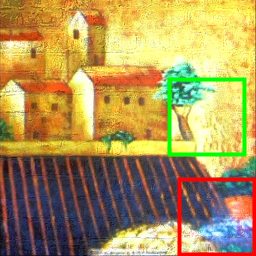}
    \includegraphics[width=\linewidth]{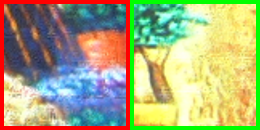}
  \end{minipage}
  \begin{minipage}[t]{0.088\textwidth}
    \centering
    \includegraphics[width=\linewidth]{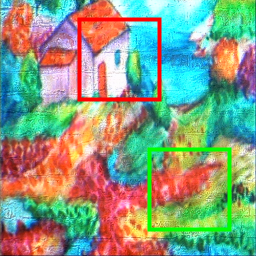}
    \includegraphics[width=\linewidth]{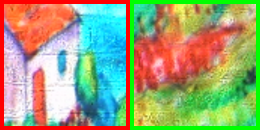}
    \includegraphics[width=\linewidth]{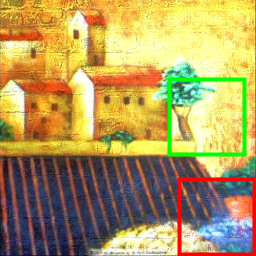}
    \includegraphics[width=\linewidth]{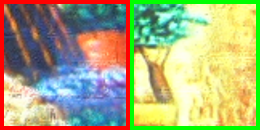}
  \end{minipage}
  \begin{minipage}[t]{0.088\textwidth}
    \centering
    \includegraphics[width=\linewidth]{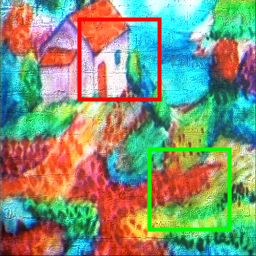}
    \includegraphics[width=\linewidth]{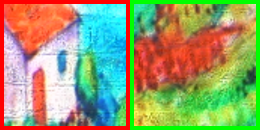}
    \includegraphics[width=\linewidth]{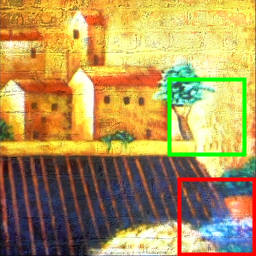}
    \includegraphics[width=\linewidth]{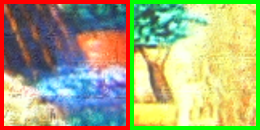}
  \end{minipage}
  \begin{minipage}[t]{0.088\textwidth}
    \centering
    \includegraphics[width=\linewidth]{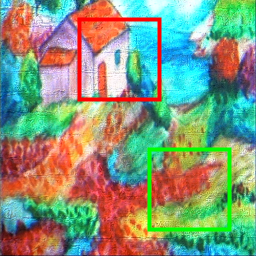}
    \includegraphics[width=\linewidth]{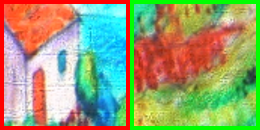}
    \includegraphics[width=\linewidth]{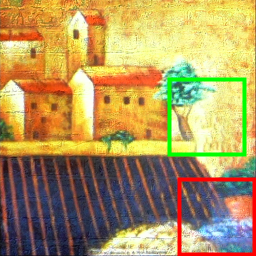}
    \includegraphics[width=\linewidth]{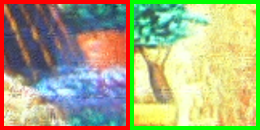}
  \end{minipage}
  \begin{minipage}[t]{0.088\textwidth}
    \centering
    \includegraphics[width=\linewidth]{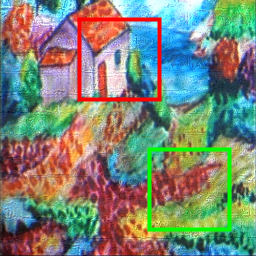}
    \includegraphics[width=\linewidth]{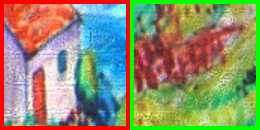}
    \includegraphics[width=\linewidth]{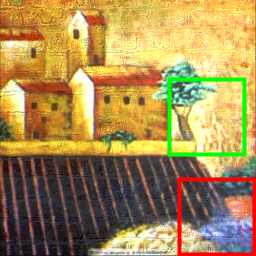}
    \includegraphics[width=\linewidth]{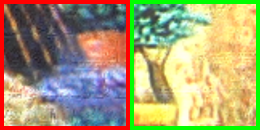}
  \end{minipage}
  \begin{minipage}[t]{0.088\textwidth}
    \centering
    \includegraphics[width=\linewidth]{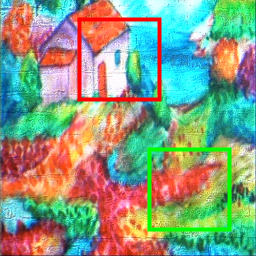}
    \includegraphics[width=\linewidth]{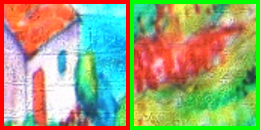}
    \includegraphics[width=\linewidth]{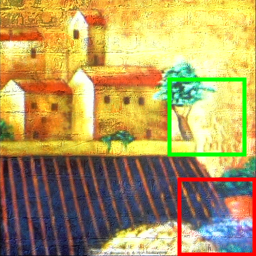}
    \includegraphics[width=\linewidth]{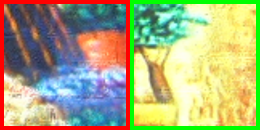}
    
  \end{minipage}
  \begin{minipage}[t]{0.088\textwidth}
    \centering
    \includegraphics[width=\linewidth]{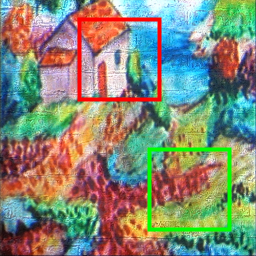}
    \includegraphics[width=\linewidth]{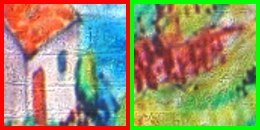}
    \includegraphics[width=\linewidth]{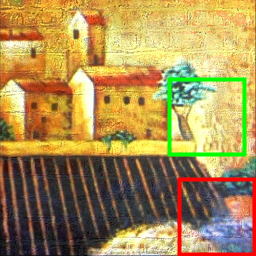}
    \includegraphics[width=\linewidth]{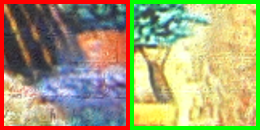}
  \end{minipage}
  \begin{minipage}[t]{0.088\textwidth}
    \centering
    \includegraphics[width=\linewidth]{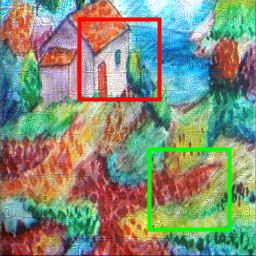}
    \includegraphics[width=\linewidth]{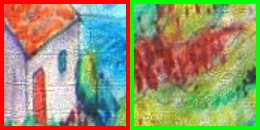}
    \includegraphics[width=\linewidth]{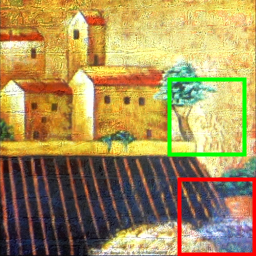}
    \includegraphics[width=\linewidth]{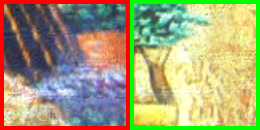}
  \end{minipage}
  \begin{minipage}[t]{0.088\textwidth}
    \centering
    \includegraphics[width=\linewidth]{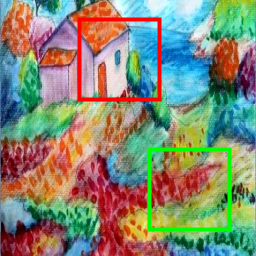}
    \includegraphics[width=\linewidth]{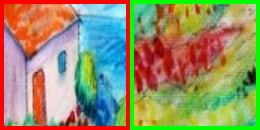}
    \includegraphics[width=\linewidth]{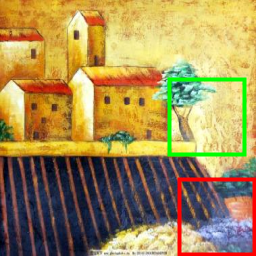}
    \includegraphics[width=\linewidth]{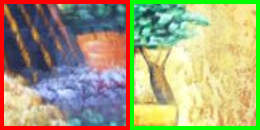}
  \end{minipage}
  \vskip 0.2cm
  \begin{minipage}[t]{0.088\textwidth}
    \centering
    \includegraphics[width=\linewidth]{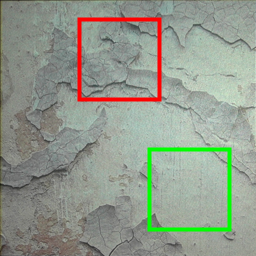}
    \includegraphics[width=\linewidth]{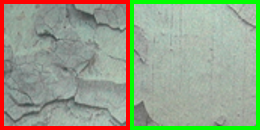}
    \includegraphics[width=\linewidth]{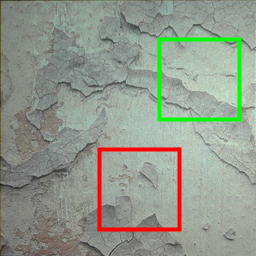}
    \includegraphics[width=\linewidth]{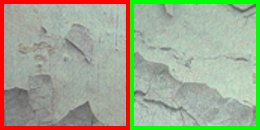}
  \end{minipage}
  \begin{minipage}[t]{0.088\textwidth}
    \centering
    \includegraphics[width=\linewidth]{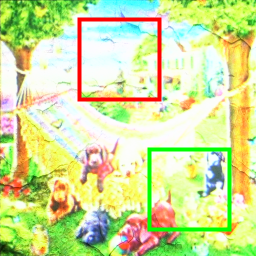}
    \includegraphics[width=\linewidth]{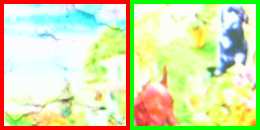}
    \includegraphics[width=\linewidth]{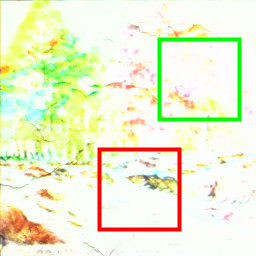}
    \includegraphics[width=\linewidth]{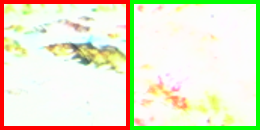}
  \end{minipage}
  \begin{minipage}[t]{0.088\textwidth}
    \centering
    \includegraphics[width=\linewidth]{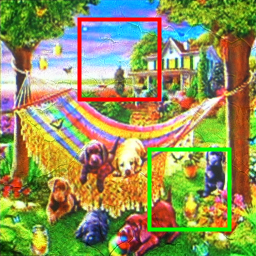}
    \includegraphics[width=\linewidth]{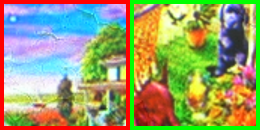}
    \includegraphics[width=\linewidth]{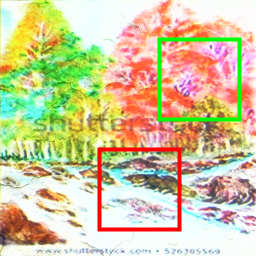}
    \includegraphics[width=\linewidth]{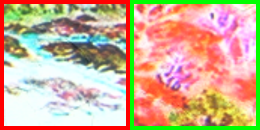}
  \end{minipage}
  \begin{minipage}[t]{0.088\textwidth}
    \centering
    \includegraphics[width=\linewidth]{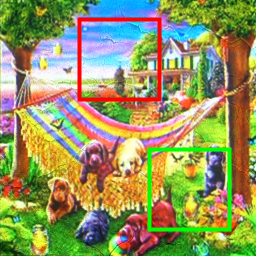}
    \includegraphics[width=\linewidth]{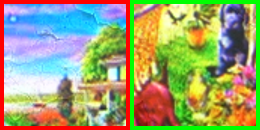}
    \includegraphics[width=\linewidth]{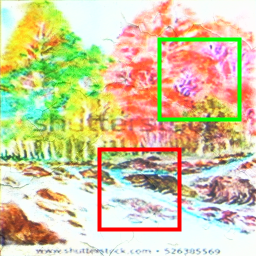}
    \includegraphics[width=\linewidth]{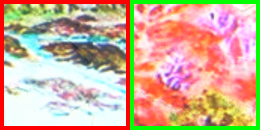}
  \end{minipage}
  \begin{minipage}[t]{0.088\textwidth}
    \centering
    \includegraphics[width=\linewidth]{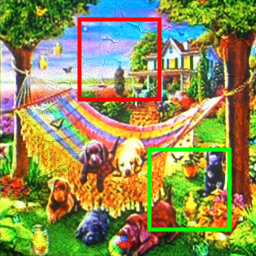}
    \includegraphics[width=\linewidth]{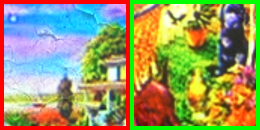}
    \includegraphics[width=\linewidth]{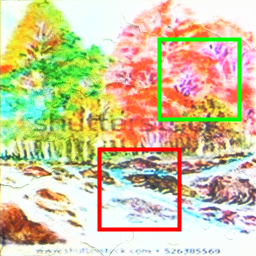}
    \includegraphics[width=\linewidth]{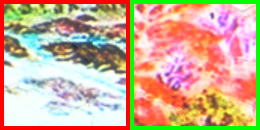}
  \end{minipage}
  \begin{minipage}[t]{0.088\textwidth}
    \centering
    \includegraphics[width=\linewidth]{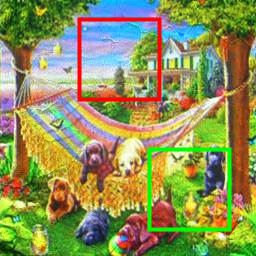}
    \includegraphics[width=\linewidth]{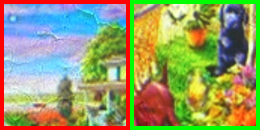}
    \includegraphics[width=\linewidth]{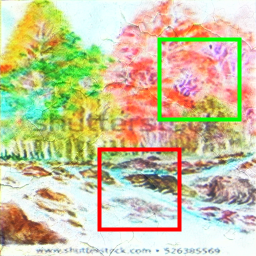}
    \includegraphics[width=\linewidth]{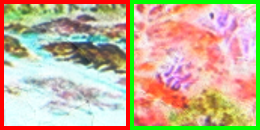}
    
  \end{minipage}
  \begin{minipage}[t]{0.088\textwidth}
    \centering
    \includegraphics[width=\linewidth]{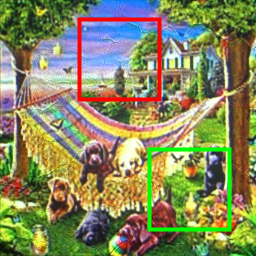}
    \includegraphics[width=\linewidth]{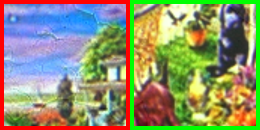}
    \includegraphics[width=\linewidth]{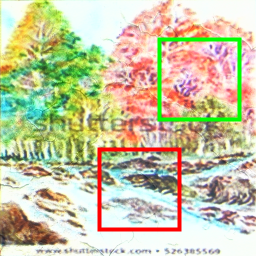}
    \includegraphics[width=\linewidth]{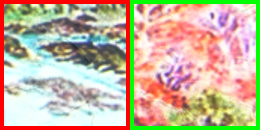}
  \end{minipage}
  \begin{minipage}[t]{0.088\textwidth}
    \centering
    \includegraphics[width=\linewidth]{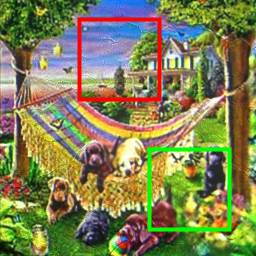}
    \includegraphics[width=\linewidth]{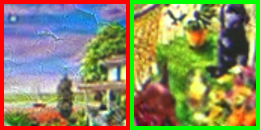}
    \includegraphics[width=\linewidth]{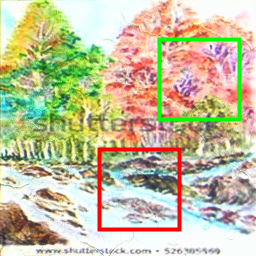}
    \includegraphics[width=\linewidth]{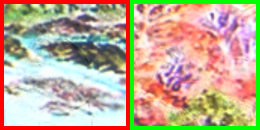}
  \end{minipage}
  \begin{minipage}[t]{0.088\textwidth}
    \centering
    \includegraphics[width=\linewidth]{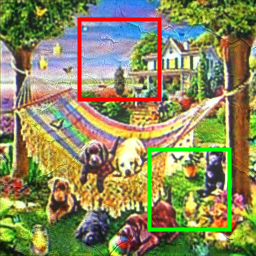}
    \includegraphics[width=\linewidth]{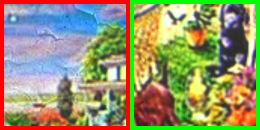}
    \includegraphics[width=\linewidth]{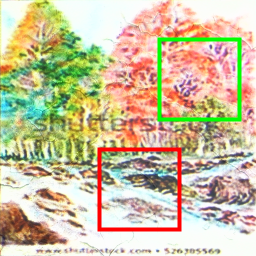}
    \includegraphics[width=\linewidth]{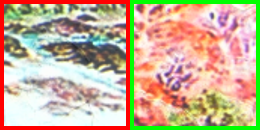}
  \end{minipage}
  \begin{minipage}[t]{0.088\textwidth}
    \centering
    \includegraphics[width=\linewidth]{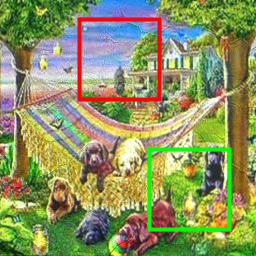}
    \includegraphics[width=\linewidth]{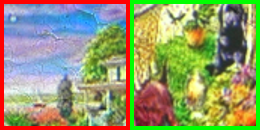}
    \includegraphics[width=\linewidth]{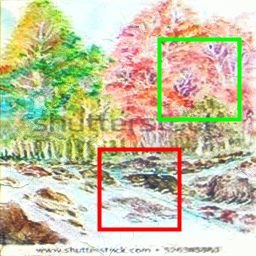}
    \includegraphics[width=\linewidth]{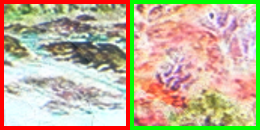}
  \end{minipage}
  \begin{minipage}[t]{0.088\textwidth}
    \centering
    \includegraphics[width=\linewidth]{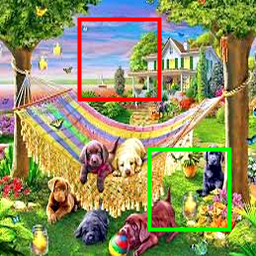}
    \includegraphics[width=\linewidth]{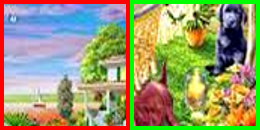}
    \includegraphics[width=\linewidth]{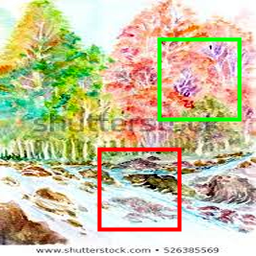}
    \includegraphics[width=\linewidth]{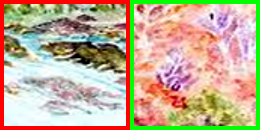}
  \end{minipage}
  \vskip 0.2cm
  \begin{minipage}[t]{0.088\textwidth}
    \centering
    \includegraphics[width=\linewidth]{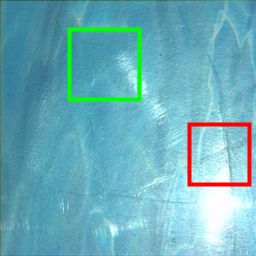}\\
    \includegraphics[width=\linewidth]{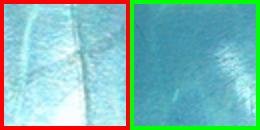}\\
    \includegraphics[width=\linewidth]{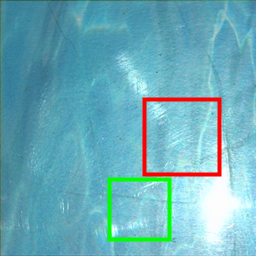}\\
    \includegraphics[width=\linewidth]{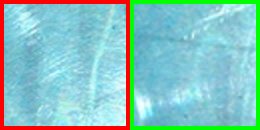}\\
  \end{minipage}
  \begin{minipage}[t]{0.088\textwidth}
    \centering
    \includegraphics[width=\linewidth]{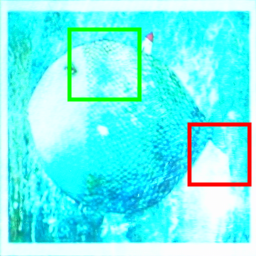}
    \includegraphics[width=\linewidth]{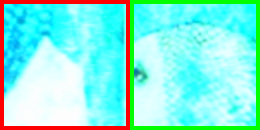}
    \includegraphics[width=\linewidth]{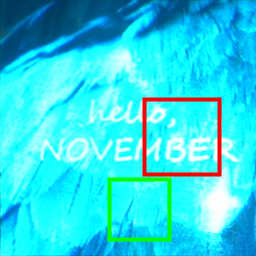}
    \includegraphics[width=\linewidth]{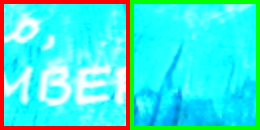}
  \end{minipage}
  \begin{minipage}[t]{0.088\textwidth}
    \centering
    \includegraphics[width=\linewidth]{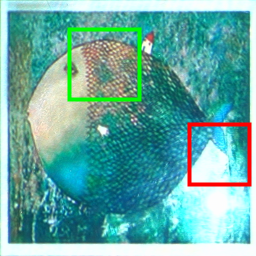}
    \includegraphics[width=\linewidth]{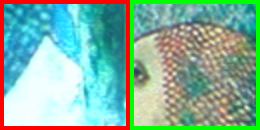}
    \includegraphics[width=\linewidth]{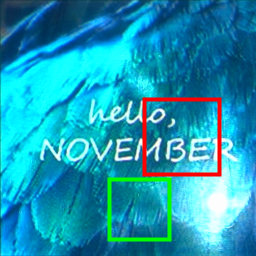}
    \includegraphics[width=\linewidth]{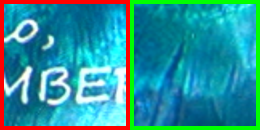}
  \end{minipage}
  \begin{minipage}[t]{0.088\textwidth}
    \centering
    \includegraphics[width=\linewidth]{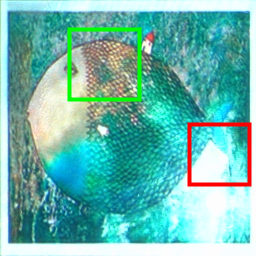}
    \includegraphics[width=\linewidth]{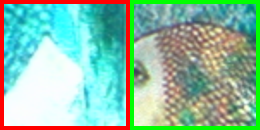}
    \includegraphics[width=\linewidth]{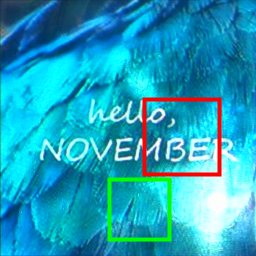}
    \includegraphics[width=\linewidth]{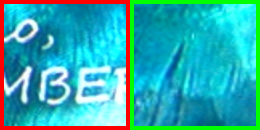}
  \end{minipage}
  \begin{minipage}[t]{0.088\textwidth}
    \centering
    \includegraphics[width=\linewidth]{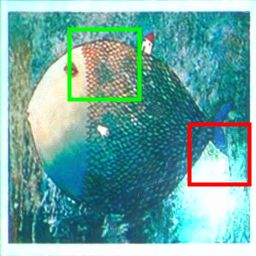}
    \includegraphics[width=\linewidth]{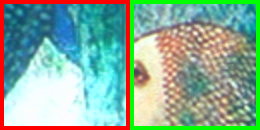}
    \includegraphics[width=\linewidth]{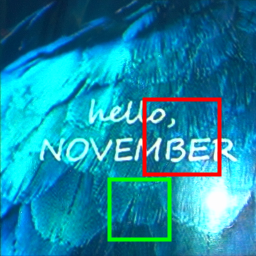}
    \includegraphics[width=\linewidth]{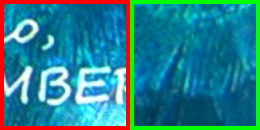}
  \end{minipage}
  \begin{minipage}[t]{0.088\textwidth}
    \centering
    \includegraphics[width=\linewidth]{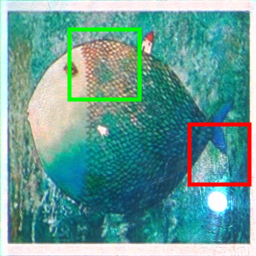}
    \includegraphics[width=\linewidth]{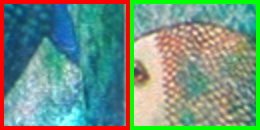}
    \includegraphics[width=\linewidth]{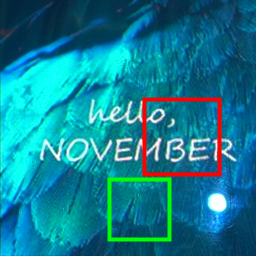}
    \includegraphics[width=\linewidth]{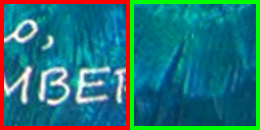}
  \end{minipage}
  \begin{minipage}[t]{0.088\textwidth}
    \centering
    \includegraphics[width=\linewidth]{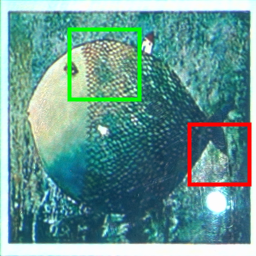}
    \includegraphics[width=\linewidth]{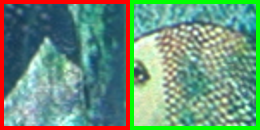}
    \includegraphics[width=\linewidth]{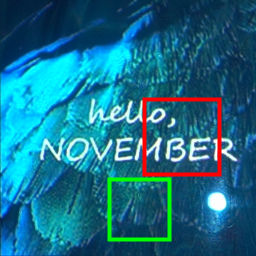}
    \includegraphics[width=\linewidth]{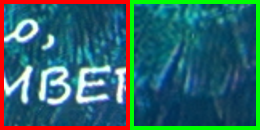}
  \end{minipage}
  \begin{minipage}[t]{0.088\textwidth}
    \centering
    \includegraphics[width=\linewidth]{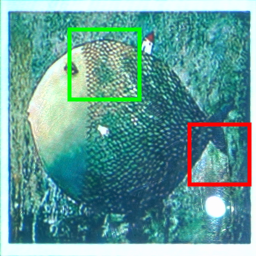}
    \includegraphics[width=\linewidth]{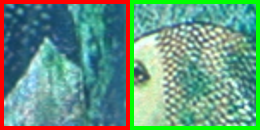}
    \includegraphics[width=\linewidth]{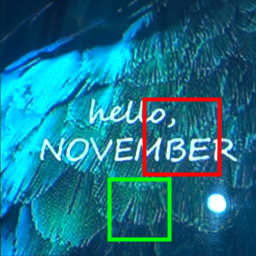}
    \includegraphics[width=\linewidth]{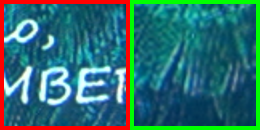}
  \end{minipage}
  \begin{minipage}[t]{0.088\textwidth}
    \centering
    \includegraphics[width=\linewidth]{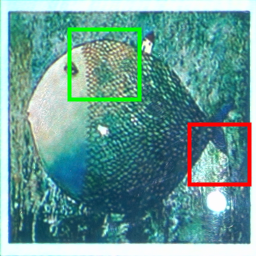}
    \includegraphics[width=\linewidth]{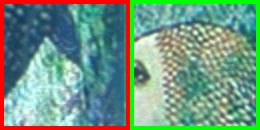}
    \includegraphics[width=\linewidth]{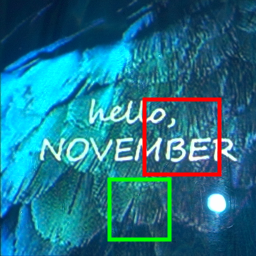}
    \includegraphics[width=\linewidth]{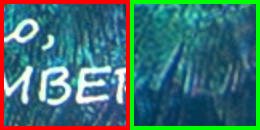}
  \end{minipage}
  \begin{minipage}[t]{0.088\textwidth}
    \centering
    \includegraphics[width=\linewidth]{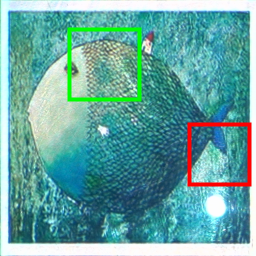}
    \includegraphics[width=\linewidth]{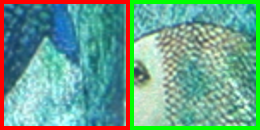}
    \includegraphics[width=\linewidth]{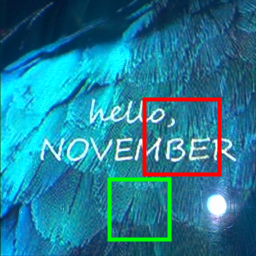}
    \includegraphics[width=\linewidth]{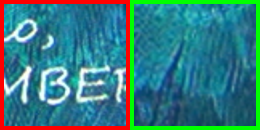}
  \end{minipage}
  \begin{minipage}[t]{0.088\textwidth}
    \centering
    \includegraphics[width=\linewidth]{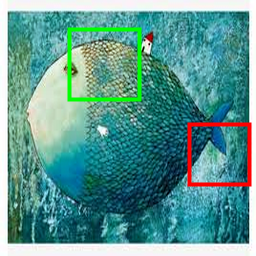}
    \includegraphics[width=\linewidth]{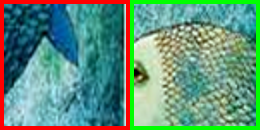}
    \includegraphics[width=\linewidth]{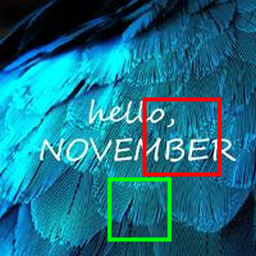}
    \includegraphics[width=\linewidth]{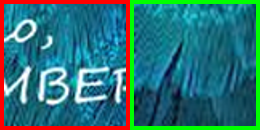}
  \end{minipage}
  \caption{Qualitative results of real-world projection. From left to right: captured surface, display results of uncompensated, pre-trained CompenNet, Pre-trained CompenNeSt, Pre-trained PANet, our pre-trained model, fine-tuned CompenNet, fine-tuned CompenNeSt, fine-tuned PANet, our fine-tuned model, and desired images (ground truth).}
  \label{fig:realprojection}
  \end{figure*}
    \subsection{Ablation study}
    To comprehensively evaluate the effectiveness of DiffPC, we conduct experiments under diverse photometry-aware conditions and with different condition-injection strategies. In these experiments, the same setups as in \cref{sec:pubexp} are employed for both model pre-training and testing, and then we compare the performance of each pre-trained model on the public datasets.
    \subsubsection{Photometry-aware condition}
    To investigate the impact of the photometry-aware condition on compensation performance, we conducted experiments using different numbers of reference images. Specifically, we compared using a single image ($w/ s1$) and three images ($w/ s3$) as photometry-aware conditions to the noise estimation network. As reported in \cref{tab:surface}, employing three photometry-aware images consistently improves performance across all evaluation metrics. Compared to the single-image condition, $w/ s3$ achieves higher SSIM and PSNR scores and lower FID, LPIPS {and $\Delta E$} values. These results validate the effectiveness of our photometry-aware conditioning strategy. 
    They highlight that richer photometry-aware information enhances the model’s ability to capture surface property, environment lighting, and projector property, thereby improving photometric compensation performance. 

    \begin{table}[!htb]
        \centering
        \small
        \caption{The average results of pre-trained models using different photometry-aware conditions on two public datasets.}
        \label{tab:surface}
        \begin{tabular}{lccccc}
         \toprule
         & PSNR $\uparrow$ & SSIM $\uparrow$ & FID $\downarrow$ & LPIPS $\downarrow$ & $\Delta E$ $\downarrow$\\
        \midrule
         Uncompensated  &  12.2762 & 0.5088 & 242.8983 & 0.4400 & 22.4814\\
           DiffPC-pre($w/ s1$)  &  18.4060	& 0.6816	& 88.5230	& 0.2895	& 11.2913\\ 
           DiffPC-pre($w/ s3$) &  {18.9552}	& 0.6996	& 76.0005 &	0.2606	& 10.3067\\
      \bottomrule
      \end{tabular}
    \end{table}

    \subsubsection{Architecture of the noise estimation network}
    Our noise estimation network is designed to integrate features from both the content condition and the photometry-aware condition into the encoder and decoder. By simultaneously incorporating these two sources of information at multiple stages of the network, our model is able to more accurately capture the interactions between the projector content and the projection environment.

    To validate the effectiveness of our architecture, we constructed comparative models using alternative feature fusion strategies, including fusion in only the encoder or decoder, as well as models that use concatenation instead of subtractive operations between the two conditions. As listed in \cref{tab:arch}, fusing features from two conditional inputs in both the encoder and decoder significantly improves performance across all evaluation metrics. Additionally, applying subtractive operations between the content and photometry-aware conditions yields further gains. 
    This is because, in our task, noise arises from factors such as illumination, surface texture, reflection, and ambient lighting. Moreover, due to the limitations of the hardware devices, the distribution of this noise is also content-dependent. As a result, the photometry-aware and content conditions can provide the U-Net with effective guidance on the noise distribution. Feature fusion in the encoding stage enables the main U-Net branch to capture condition-guided details, while fusion in the decoding stage provides high-level semantic information from these conditions. Finally, compared with concatenation, the subtractive operation emphasizes the differences between the two conditions, which better reflects the physical process by which distortions are introduced during projection. Therefore, by combining these two strategies, our architecture achieves the best overall performance.


    \begin{table}[!htb]
        \centering
        \small
        \caption{The average results of denoising networks with different architectures on three public datasets. (Sub: subtraction, Cat: concatenation, E: used in encoder, D: used in decoder)}
        \label{tab:arch}
        \begin{tabular}{lccccc}
         \toprule
           & PSNR $\uparrow$ & SSIM $\uparrow$ & FID $\downarrow$ & LPIPS $\downarrow$ & $\Delta E$ $\downarrow$\\
         \midrule
           Uncompensated  &  12.2762	& 0.5088 &	242.8983 &	0.4400 & 22.4814 \\
           Sub+CatE  &  18.6509	   & 0.6795	  & 81.5824	& 0.2707	& 10.8208\\
           Sub+CatD  &  18.8135	& 0.6848  & 86.4994	& 0.2721	& 10.5527\\
           Sub+AddD  &  18.6843	& 0.6751	& 92.0405 & 	0.2876	& 10.8687\\
           Cat+CatD  &  17.9981	& 0.6594	& 87.5081  &	0.2977	& 11.4618 \\ 
           Cat+CatED  & 18.6476	& 0.6856	& 79.3226  &	0.2739 & 11.0013 \\
           Sub+CatED(ours)  & 18.9552	& 0.6996	& 76.0005  &	0.2606 & 10.3067 \\
       \bottomrule
      \end{tabular}
    \end{table}
    
    
    \section{Conclusion and discussion}
    In this paper, we proposed a diffusion-based photometric compensation method that effectively addresses color distortions in static projector–camera systems. By reformulating the photometric compensation task as a denoising problem, our approach leverages a diffusion model for progressive compensation. In addition, to learn the distribution of the noise caused by the physical process of projection, we designed a noise estimation network that is guided by the photometry-aware and content conditions. Extensive experiments demonstrate that our method achieves superior generalization across diverse surfaces and illumination conditions, while delivering high perceptual quality in the compensation images.

    Despite its effectiveness, our method has some limitations.
    As a data-driven learning-based method, the performance of our method depends on the training data distribution. {Due to the difficulty of collecting real pairs, we followed \cite{Huang_CVPR_2019} to use surrogate training data, which exhibits a data distribution gap compared to real environments. Resolving this discrepancy is a challenge for future work.}
    Moreover, although efficient blocks are used in the noise estimation network, the diffusion-based sampling speed still limits real-time applications. {Applying advanced acceleration techniques is a key priority for our future research, which would enable real-time compensation for moving projection surfaces.}
    
    

\acknowledgments{
Wang was supported in part by the Nature Science Foundation of China (62202135). Huang was supported in part by the Nature Science
Foundation of China (62302401).}

\bibliographystyle{abbrv-doi}

\bibliography{cmp}
\end{document}